\documentclass[pra,twocolumn,floatfix,superscriptaddress,longbibliography,notitlepage]{revtex4-2}
\usepackage{amssymb,amsmath,amsthm,color,graphicx,times,graphicx}
\usepackage{caption}
\usepackage{ragged2e}
\DeclareCaptionJustification{justified}{\justifying}
\usepackage{hyperref}
\usepackage{braket}
\usepackage{graphicx}
\usepackage{dcolumn}
\usepackage{appendix}
\usepackage{subcaption}
\usepackage{bm}

\providecommand{\openone}{\leavevmode\hbox{\small1\kern-4.3pt\normalsize1}}

 \usepackage{orcidlink}

\theoremstyle{plain}

\theoremstyle{definition}

\begin{document}
\title{Hyperon-antihyperon system in electron-positron annihilation as quantum probes for temperature estimation with local and global dephasing}

\author{Anass Hminat \orcidlink{0009-0007-3677-3952}}\affiliation{LPHE-Modeling and Simulation, Faculty of Sciences, Mohammed V University in Rabat, Rabat, Morocco.}
\author{Abdallah Slaoui \orcidlink{0000-0002-5284-3240}}\email{abdallah.slaoui@um5s.net.ma}\affiliation{LPHE-Modeling and Simulation, Faculty of Sciences, Mohammed V University in Rabat, Rabat, Morocco.}\affiliation{Centre of Physics and Mathematics, CPM, Faculty of Sciences, Mohammed V University in Rabat, Rabat, Morocco.}
\author{Rachid Ahl Laamara \orcidlink{0000-0002-8254-9085}}\affiliation{LPHE-Modeling and Simulation, Faculty of Sciences, Mohammed V University in Rabat, Rabat, Morocco.}\affiliation{Centre of Physics and Mathematics, CPM, Faculty of Sciences, Mohammed V University in Rabat, Rabat, Morocco.}
\author{Hichem Eleuch \orcidlink{0000-0002-4596-137X}}\affiliation{Univ. Polytechnique Hauts-de-France, LAMIH, CNRS, UMR 8201, F-59313 Valenciennes, France.}

\begin{abstract}
We investigate quantum thermometry in Ohmic-type reservoirs using two-qubit probes within an exactly solvable pure-dephasing framework. By analyzing the individual variance associated with temperature estimation, we identify optimal regimes governed by the Ohmicity parameter $s$, the deviation angle $\theta$, and the decay coefficients $\alpha$ and $\beta$, thereby determining the conditions that minimize estimation errors. The Quantum Fisher Information (QFI) exhibits pronounced maxima at finite interaction times, especially in sub-Ohmic and Ohmic environments at low temperatures, whereas super-Ohmic reservoirs flatten the QFI peak and shift the optimal sensitivity toward higher temperatures. Consistently, the quantum signal-to-noise ratio (QSNR) is suppressed at low temperatures, increases with thermal excitation, and saturates in the high-temperature regime, where the influence of spectral details becomes negligible. A comparative study of mutual and local estimation strategies shows that common-bath configurations, particularly for $\Sigma^+$ and $\Sigma^0$ probes, outperform local baths at short interaction times due to bath-induced correlations, while local environments become advantageous at longer times. The analysis further reveals finite optimal values of both the interaction time $t_{\rm opt}$ and the temperature $T_{\rm opt}$, as well as a strong reduction of the variance with increasing measurement number in the low-temperature regime. In addition, our study of hyperon-antihyperon channels ($\Lambda$, $\Sigma^+$, $\Sigma^0$, $\Xi^-$, $\Xi^0$) shows that entanglement and quantum discord remain remarkably robust over broad angular domains, whereas steering and Bell nonlocality are confined to narrower regions. Overall, the interplay between spectral structure, particle-dependent parameters, and estimation strategy provides valuable insight for designing high-precision quantum thermometers and exploiting quantum correlations in metrological applications, particularly in low-temperature settings.
\end{abstract}

\par
\vspace{0.25cm}

\date{\today}

\maketitle

\section{introduction}
Thermometry consists in estimating the temperature of a large system by means of a smaller auxiliary device, called a probe \cite{1}. In the usual classical picture, this procedure relies on the zeroth law of thermodynamics: the sample is assumed to possess a much larger heat capacity than the probe \cite{2,3}. Once both are brought into contact, energy exchange leads them to reach thermal equilibrium \cite{4}. Because of its much larger heat capacity, the sample is only weakly affected, while the probe eventually attains the same temperature as the system and can then be used as a thermometer \cite{5}. In realistic situations, however, the heat capacity of the sample is never truly infinite, so any temperature reading inevitably causes some disturbance. Moreover, for quantum systems with a finite energy gap, accurate thermometry becomes impossible below a certain temperature threshold \cite{6}. Quantum probes offer an alternative and potentially more powerful route to temperature estimation. In recent years, the use of quantum states and quantum measurements for thermometry has attracted significant attention, especially because coherence and interference can improve performance \cite{gx1,gx3}. Such probes may allow one to infer temperature while only minimally perturbing the system under study \cite{9}. Among the available options, single-qubit probes are particularly appealing since they represent the simplest quantum systems capable of extracting information from the environment \cite{10}. It is also interesting to consider estimation strategies that do not rely on thermalization or on the zeroth-law picture of energy exchange between the probe and the sample \cite{11}. In this work, we focus on the dephasing dynamics of a qubit as an effective way to estimate the temperature of its surroundings \cite{12}. This approach is intrinsically quantum, since it exploits the sensitivity of quantum systems to decoherence and does not require the probe to reach thermal equilibrium with the system \cite{13}. High-energy colliders offer an important environment for testing quantum entanglement and nonlocality \cite{40}. Thanks to major advances in accelerator and detector performance, experiments now collect data with sufficient precision to make quantum correlations accessible in high-energy reactions \cite{41}. In recent years, such effects have been studied in several elementary-particle systems, including top-quark pairs at the Large Hadron Collider, lepton pairs, and gauge bosons produced in Higgs decays \cite{42,43,44,45}. Compared with these systems, hadronic final states have a longer history in this context \cite{46}. More recently, hyperon systems have attracted considerable attention because the weak decay of a hyperon acts as a built-in polarimeter, allowing spin observables such as polarization and spin correlations to be extracted experimentally. With the upgraded Beijing Spectrometer III (BESIII) at the Beijing Electron-Positron Collider \cite{61,62}, hyperon-antihyperon production in $e^+e^-$ annihilation has become a promising platform for investigating quantum correlations.

Quantum entanglement has become an increasingly active topic in high-energy physics\cite{47,gx2,gx4}, especially in studies of hadronic final states produced in proton-proton collisions , where quantum effects can be probed at very short distances \cite{48}. Several systems have emerged as particularly interesting testbeds for entanglement and CP-violation studies, including neutral kaons, neutral $B$ mesons, and positronium \cite{50,51,52}. More recently, entanglement in top-quark pair production at the LHC has drawn significant attention, together with proposals for observing Bell inequality violation \cite{53}. These developments have stimulated further work on entanglement in a variety of systems, such as top-quark pairs, hyperon pairs, and gauge bosons produced either in Higgs decays or through direct production \cite{54,55,56}. Experimental evidence for entanglement has already been reported at $\sqrt{s}=13~\mathrm{TeV}$, and Bell-inequality violation has also been discussed in $B$-meson decays at LHCb and Belle II \cite{58}.

Understanding baryon production also requires the study of effective transition form factors, particularly in processes such as electron--positron annihilation \cite{57}, where baryons may be produced through a virtual photon or via resonance states. Within the helicity formalism, these form factors encode the spin and polarization structure of the process and make it possible to compute helicity transition amplitudes as well as polarization-correlation coefficients in two-baryon systems \cite{59}. This approach not only provides insight into baryon structure, but also offers a natural framework for investigating CP violation by comparing baryon and antibaryon production channels \cite{61}, thereby shedding light on matter--antimatter asymmetry. Taken together, these ingredients provide a powerful framework for studying baryon dynamics and fundamental symmetries in particle physics \cite{62}.

The structure of the paper is as follows. In Sec.~II, we give a concise overview of the key elements of quantum estimation theory that will be used in the rest of the work, then we present the two-qubit density operator describing the hyperon--antihyperon system produced in $e^+e^-$ annihilation. We then introduces the physical setting of a two-qubit system coupled to a bosonic thermal reservoir, and examines how the dynamics changes when the qubits interact either with a shared environment or with two separate but identical baths. In Sec.~II, we analyze the behavior of the minimum variance of the temperature for different internal parameters in order to determine the optimal parameter configuration that minimizes the estimation error, we then estimate the temperature by calculating the quantum Fisher information and the quantum signal noise. We perform a comparative analysis between mutual (collective) and local bosonic coupling in order to assess their impact on temperature estimation and to identify the optimal configuration for the hyperons under consideration. Subsequently, we determine the optimal interaction time and temperature that maximize the precision of the temperature measurement. In the section IV,  we discuss the hierarchy of quantum correlations. Finally, in the last section V we provides the concluding remarks of the paper.

\section{Quantum thermometry protocol}
We study a pure-dephasing scenario in which two spin-$1/2$ hyperon and an antihyperon are coupled to a bosonic environment at temperature $T$. The reservoir is characterised by an Ohmic-class spectral density, a choice that permits an exact analytic treatment. Here the hyperon and antihyperon are employed as quantum probes to estimate the temperature T of the reservoir, rather than assuming prior knowledge of the bath and focusing solely on decoherence effects. This framework enables us to examine the influence of quantum correlations, as well as the impact on the precision of temperature estimation. In the following section, we will present the physical model and the nature of the generated hyperons and antihyperons resulting from the scattering process. We will then analyze the dynamics induced by the ohmic dephasing of the reservoir, considering both mutual and local dephasing. Finally, we will introduce the mathematical tools of quantum estimation theory in order to perform a thermometric study of our probe and to evaluate the effect of ohmicity on the measurement of the temperature generated by the reservoir.

\subsection{The physical model}

Hyperon-antihyperon pairs can be created in electron--positron annihilation either through an intermediate virtual photon,
$
e^+e^- \to \gamma^* \to Y\bar{Y},
$
or via vector charmonium decays such as
$e^+e^- \to J/\psi \to Y\bar{Y}$. Here, $Y$ represents a ground-state octet hyperon, namely $\Lambda$, $\Sigma^+$, $\Xi^-$, or $\Xi^0$. At BESIII, large data samples of vector charmonia such as $J/\psi$ and $\psi(2S)$ have been accumulated, and these resonances can decay into hyperon--antihyperon final states. Since both particles have spin $1/2$, the pair forms a massive two-qubit system. In the center-of-mass frame, momentum conservation forces the hyperon and antihyperon to emerge in opposite directions, so their momenta are back-to-back. The spin–correlation matrix can be written as
\begin{equation}
\mathbb{B}_{ij} =
\begin{pmatrix}
 \frac{\sin^2\vartheta }{1 + \alpha_{\psi}\cos^2\vartheta}& 0 &  \frac{\gamma_{\psi}\sin\vartheta\cos\vartheta}{1 + \alpha_{\psi}\cos^2\vartheta} \\
0 & - \frac{\alpha_{\psi}\sin^2\vartheta }{1 + \alpha_{\psi}\cos^2\vartheta}& 0 \\ \frac{\gamma_{\psi}\sin\vartheta\cos\vartheta}{1 + \alpha_{\psi}\cos^2\vartheta}
 & 0 & \frac{\alpha_{\psi} + \cos^2\vartheta}{1 + \alpha_{\psi}\cos^2\vartheta} 
\end{pmatrix},
\end{equation}
where $\alpha_{\psi} \in [-1,1]$ denotes the decay asymmetry parameter, 
$\Delta\Phi \in (-\pi,\pi]$ represents the relative phase, and $\vartheta$ 
is the scattering angle between the incoming electron beam and the outgoing hyperon.The quantities $\beta_{\psi}$ and $\gamma_{\psi}$ are defined as
\begin{equation}
\beta_{\psi} = \sqrt{1 - \alpha_{\psi}^2}\,\sin(\Delta\Phi), 
\qquad
\gamma_{\psi} = \sqrt{1 - \alpha_{\psi}^2}\,\cos(\Delta\Phi).
\end{equation}

\begin{figure}[t] 
        \includegraphics[width=7.8cm]{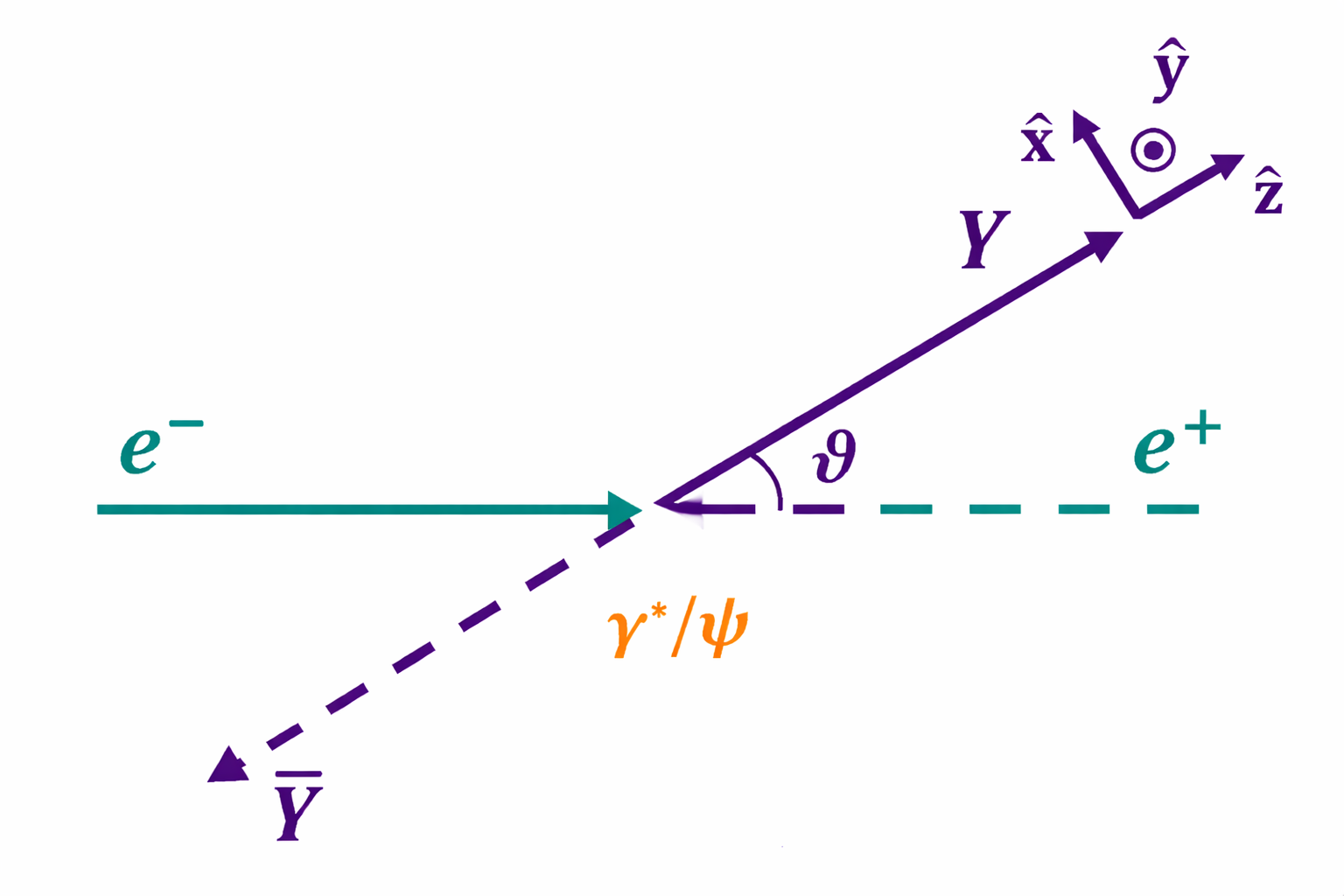}
      \caption{The coordinate system adopted in the analysis is defined by the orthogonal directions $\{\hat{x},\hat{y},\hat{z}\}$ in the rest frame of $Y$, and equivalently in that of $\overline{Y}$.}
    \label{fig4}
\end{figure}

These parameters are experimentally determined, for instance by the BESIII collaboration, 
with representative values reported in Table~I. The vectors $\mathbf{B}^{\pm}$ 
and the matrix $C_{ij}$ are defined in the respective rest frames of the hyperon 
and antihyperon.

Next, we introduce an orthonormal coordinate system defined by
\begin{equation}
\hat{\mathbf{y}} = \frac{\hat{\mathbf{p}}_{Y} \times \hat{\mathbf{p}}_{e}}
{\left|\hat{\mathbf{p}}_{Y} \times \hat{\mathbf{p}}_{e}\right|}, 
\quad
\hat{\mathbf{z}} = \hat{\mathbf{p}}_{Y}, 
\quad
\hat{\mathbf{x}} = \hat{\mathbf{y}} \times \hat{\mathbf{z}},
\end{equation}
which applies symmetrically to both the hyperon and the antihyperon.

As illustrated in Fig.~1, the antihyperon $\bar{Y}$ is also described in its own rest frame, but its axes are taken to coincide with those of the hyperon, namely
\[
\{\hat{x}_{\bar{Y}},\hat{y}_{\bar{Y}},\hat{z}_{\bar{Y}}\}
=
\{\hat{x},\hat{y},\hat{z}\}.
\]
Therefore leads to slightly different components of $\mathbb{B}_{ij}$. The advantage of this convention is that the rest frames of $Y$ and $\bar{Y}$ are related only by a pure boost along their directions of motion, without any additional rotation. In the rest frames of $Y$ and $\bar{Y}$.

Since quantum correlations remain invariant under local unitary transformations, one can map the bipartite density operator onto a standard two-qubit $X$-state form:
\begin{equation}
\rho^{X}_{Y\bar{Y}} = \frac{1}{4} \left(
\mathbb{I}\otimes\mathbb{I}
+ a\,\sigma_z \otimes \mathbb{I}
+ \mathbb{I} \otimes a\,\sigma_z
+ \sum_{i=1}^{3} t_i\,\sigma_i \otimes \sigma_i
\right),
\end{equation}
where the coefficients are given by
\begin{equation}
a = \frac{\beta_{\psi}\sin\vartheta\cos\vartheta}
{1 + \alpha_{\psi}\cos^2\vartheta},
\end{equation}
\begin{equation}
t_{1,2} = \frac{1 + \alpha_{\psi} \pm 
\sqrt{(1 + \alpha_{\psi}\cos 2\vartheta)^2 
- (\beta_{\psi}\sin 2\vartheta)^2}}
{2(1 + \alpha_{\psi}\cos^2\vartheta)},
\end{equation}
\begin{equation}
t_3 = \frac{-\alpha_{\psi}\sin^2\vartheta}
{1 + \alpha_{\psi}\cos^2\vartheta}.
\end{equation}

From the above expressions, it follows that the original density matrix 
and its $X$-state representation are locally unitary equivalent. 
This transformation significantly simplifies the subsequent analysis.

By explicitly expanding the Pauli operators, the $X$-state density matrix 
can be written in matrix form as
\begin{equation}
\rho^{X}_{Y\bar{Y}} = \frac{1}{4}
\begin{pmatrix}
1 + 2a + t_3 & 0 & 0 & t_1 - t_2 \\
0 & 1 - t_3 & t_1 + t_2 & 0 \\
0 & t_1 + t_2 & 1 - t_3 & 0 \\
t_1 - t_2 & 0 & 0 & 1 - 2a + t_3
\end{pmatrix}.
\end{equation}

This structure is commonly referred to as an $X$-state due to the characteristic 
pattern formed by its nonzero elements. This section briefly reviews the fundamental notions required to describe correlated quantum channels. For the sake of clarity, we focus on a bipartite system composed of two qubits, initially prepared in the state $\hat{\varrho}(0)$. We will then implement a dephasing operation to describe the dynamics of the system; this will be the focus of the next subsection, where both mutual and local dephasing scenarios will be considered.

\begin{table*}[ht]
\centering
\caption{Some parameters in $e^{+}e^{-} \rightarrow J/\psi \rightarrow Y\bar{Y}$, where $Y\bar{Y}$ is a pair of ground-state octet hyperons.}
\label{tab:hyperons}
\begin{tabular}{lcccccc}
\hline\hline
Decay mode & Branching ratio ($\times 10^{-4}$) & $\alpha_{\psi}$ & $\Delta\Phi$ (rad) & Ref. \\
\hline
$J/\psi \rightarrow \Lambda \bar{\Lambda}$ 
& $19.43 \pm 0.03 \pm 0.33$
& $0.4748 \pm 0.0022 \pm 0.0031$
& $0.7521 \pm 0.0042 \pm 0.0066$
& \cite{ref40} \\

$J/\psi \rightarrow \Sigma^{+}\bar{\Sigma}^{-}$
& $10.61 \pm 0.04 \pm 0.36$
& $-0.508 \pm 0.006 \pm 0.004$
& $-0.270 \pm 0.012 \pm 0.009$
& \cite{ref42} \\

$J/\psi \rightarrow \Sigma^{0}\bar{\Sigma}^{0}$
& $11.64 \pm 0.04 \pm 0.23$
& $-0.4133 \pm 0.0035 \pm 0.0077$
& $-0.0828 \pm 0.0000 \pm 0.0033$
& \cite{ref44} \\

$J/\psi \rightarrow \Xi^{-}\bar{\Xi}^{+}$
& $10.40 \pm 0.06 \pm 0.74$
& $0.586 \pm 0.012 \pm 0.010$
& $1.213 \pm 0.046 \pm 0.016$
& \cite{ref45} \\

$J/\psi \rightarrow \Xi^{0}\bar{\Xi}^{0}$
& $11.65 \pm 0.04 \pm 0.43$
& $0.514 \pm 0.006 \pm 0.0015$
& $1.168 \pm 0.019 \pm 0.018$
& \cite{ref47} \\
\hline\hline
\end{tabular}
\end{table*}

\subsection{Quantum dephasing in common and local baths }

\subsubsection*{Mutual environment}
If both qubits interact with the \emph{same} reservoir, the total Hamiltonian becomes
\begin{equation}
H \;=\; \tfrac{1}{2}\sum_{j=1}^2 \sigma_z^{(j)}
+\sum_k \omega_k b_k^\dagger b_k
+\sum_{j=1}^2\sum_k \sigma_z^{(j)}\bigl(g_k b_k^\dagger + g_k^* b_k\bigr),
\label{eq:common_H}
\end{equation}

Again assuming identical couplings $g_k$ for the two qubits. The index $j$ identifies the qubits, whereas $k$ runs over the bath modes. The operator $\sigma_3^{(j)}$ denotes the third Pauli matrix acting on qubit $j$, and $b_k$ and $b_k^\dagger$ are the bosonic annihilation and creation operators, respectively. After passing to the continuum limit,
\[
\sum_k g_k \;\to\; \int d\omega\, J_s(\omega,\Omega)\,[2|g(\omega)|]^{-2},
\]
and letting the composite probe--bath system evolve under the above Hamiltonian up to the dimensionless time $\tau=\Omega t$, one obtains, after tracing out the bath degrees of freedom, the reduced two-qubit dynamics in the form :
\begin{equation}
\rho_{\mathrm{CB}}(\tau,T)=V(\tau,T)\circ R(\tau)\circ \rho,
\end{equation}
where $V(\tau,T)$ is given by
\begin{equation}
V(\tau,T)=
\begin{pmatrix}
1 & e^{-\Gamma_s(\tau,T)} & e^{-\Gamma_s(\tau,T)} & e^{-4\Gamma_s(\tau,T)} \\
e^{-\Gamma_s(\tau,T)} & 1 & 1 & e^{-\Gamma_s(\tau,T)} \\
e^{-\Gamma_s(\tau,T)} & 1 & 1 & e^{-\Gamma_s(\tau,T)} \\
e^{-4\Gamma_s(\tau,T)} & e^{-\Gamma_s(\tau,T)} & e^{-\Gamma_s(\tau,T)} & 1
\end{pmatrix},
\end{equation}
and
\begin{equation}
R(\tau)=
\begin{pmatrix}
1 & e^{2if(\tau)} & e^{2if(\tau)} & 1 \\
e^{-2if(\tau)} & 1 & 1 & e^{-2if(\tau)} \\
e^{-2if(\tau)} & 1 & 1 & e^{-2if(\tau)} \\
1 & e^{2if(\tau)} & e^{2if(\tau)} & 1
\end{pmatrix}.
\end{equation}
The symbol $\circ$ denotes the Hadamard (elementwise) product, and $\rho$ is the initial two-qubit state. The decoherence function depends on both time and the dimensionless temperature $T=(\Omega\beta)^{-1}$. By measuring frequencies in units of $\Omega$, namely $\omega\to\omega/\Omega$, it can be written as
\begin{equation}
\Gamma_s(\tau,T)=\int_0^\infty d\omega\,
e^{-\omega}\,
\frac{1-\cos(\omega\tau)}{\omega^{2-s}}
\coth\!\left(\frac{\omega}{2T}\right),
\end{equation}
whose closed form is available in Ref.~\cite{18}. The phase function $f(\tau)$ is independent of temperature and is defined by
\begin{equation}
f(\tau)=\frac{1}{2}\int_0^\infty d\omega\,
J_s(\omega)\,
\frac{\omega\tau-\sin(\omega\tau)}{\omega^2}.
\end{equation}
At short times, the matrix $V(\tau,T)$ dominates the evolution, while $R(\tau)$ is responsible for generating entanglement between the qubits.
\subsubsection*{Local environment}

We now consider the situation in which each qubit is locally coupled to its own environment, with the two reservoirs being identical but uncorrelated. In this configuration, the total Hamiltonian can be expressed as
\begin{align}
H = \sum_{j=1}^{2} \Bigg\{
&\frac{\omega_j}{2} \sigma_3^{(j)}
+ \sum_{k=0}^{\infty} \omega_k \, b_k^{(j)\dagger} b_k^{(j)}
\nonumber\\
&+ \sum_{k=0}^{\infty} \sigma_3^{(j)}
\left( g_k b_k^{(j)\dagger} + g_k^* b_k^{(j)} \right)
\Bigg\}.
\end{align}

The initial state of the environment is assumed to be separable, namely
\begin{equation}
\rho_B(0) = \rho_B^{(1)} \otimes \rho_B^{(2)},
\end{equation}
where each $\rho_B^{(j)}$ corresponds to a thermal Gibbs state at a dimensionless temperature $T$.

Under these assumptions, the reduced dynamics of the two-qubit probe can be written as
\begin{equation}
\rho_{LB}(\tau, T) = W(\tau, T) \circ \rho,
\end{equation}
where $\circ$ denotes the Hadamard (element-wise) product. The matrix $W(\tau, T)$ describes the action of two independent dephasing channels and is given by
\begin{equation}
W(\tau, T) =
\begin{pmatrix}
1 & e^{-\Gamma_s(\tau,T)} & e^{-\Gamma_s(\tau,T)} & e^{-2\Gamma_s(\tau,T)} \\
e^{-\Gamma_s(\tau,T)} & 1 & e^{-2\Gamma_s(\tau,T)} & e^{-\Gamma_s(\tau,T)} \\
e^{-\Gamma_s(\tau,T)} & e^{-2\Gamma_s(\tau,T)} & 1 & e^{-\Gamma_s(\tau,T)} \\
e^{-2\Gamma_s(\tau,T)} & e^{-\Gamma_s(\tau,T)} & e^{-\Gamma_s(\tau,T)} & 1
\end{pmatrix}.
\end{equation}

\begin{figure}[t] 
        \includegraphics[width=9cm]{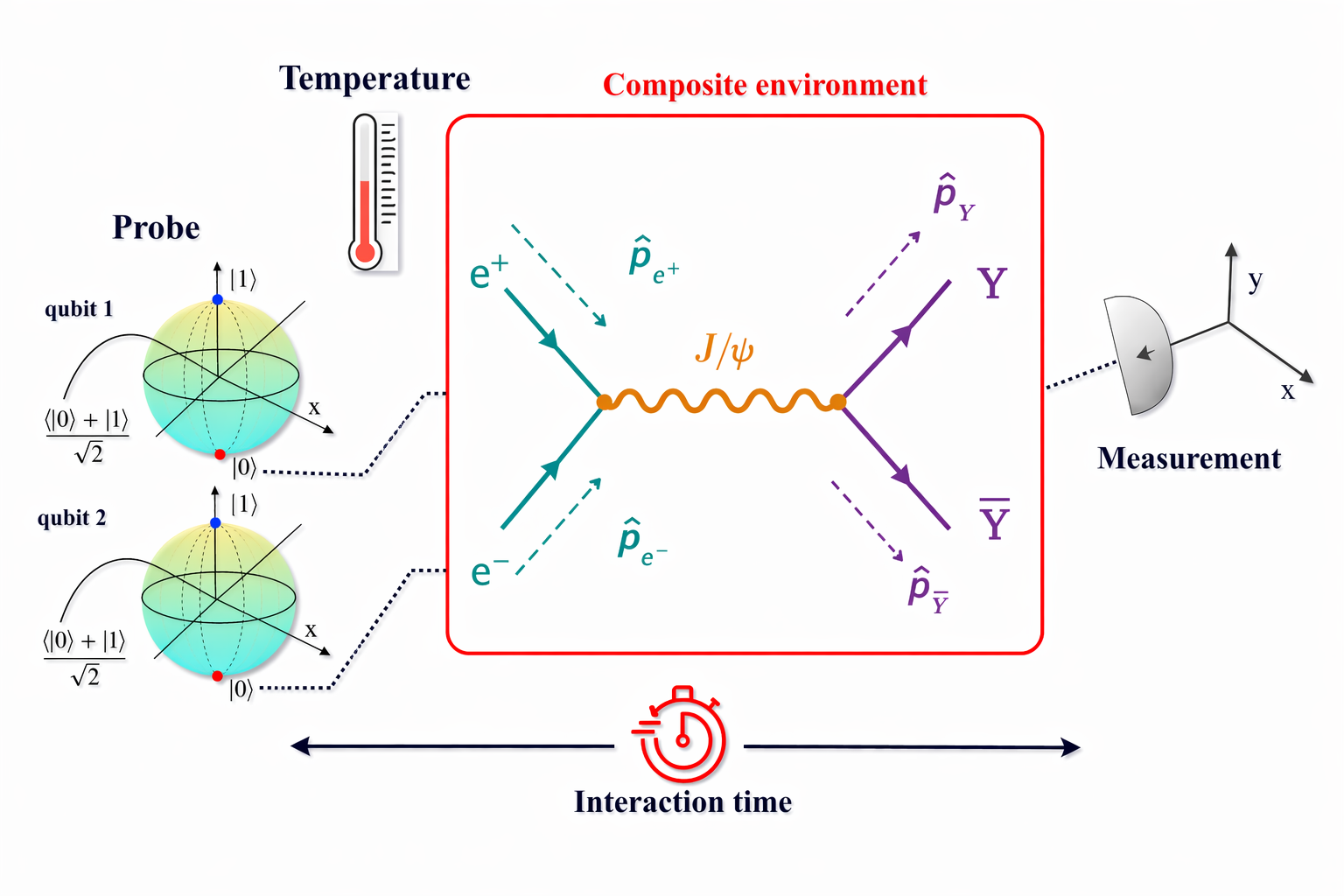}
      \caption{We consider a quantum thermometric protocol based on a single two-level system that undergoes dephasing as a result of its interaction with a structured reservoir initially at thermal equilibrium. In this picture, dephasing is illustrated by the contraction of the Bloch sphere in the interaction frame. Once the interaction has taken place, the temperature is inferred by performing a measurement along the optimal spin direction.}
    \label{fig2}
\end{figure}

In Fig.\ref{fig2}, we present a schematic of the thermometric protocol based on two probes qubits. The dephasing process is visualized as a contraction of the Bloch sphere, while the initial spin state (shown by the red point) is considered in the interaction picture. After the probe has interacted with the sample, as discussed in Sec.~4, the final state is read out by a measurement performed along the optimal spin direction.

\subsection{Quantum estimation}

In this section, we recall the mathematical tools needed to derive the quantum Fisher information matrix (QFIM). In particular, we make use of a vectorization procedure that maps a matrix into a column vector, allowing the elements of the QFIM to be written without explicitly diagonalizing the density matrix. Let $M_{n\times n}$ denote the space of $n\times n$ real or complex matrices. For any matrix $A\in M_{n\times n}$, the \emph{vec}-operator is defined as
\begin{equation}
\mathrm{vec}[A] = (a_{11},\ldots,a_{n1},a_{12},\ldots,a_{n2},\ldots,a_{1n},\ldots,a_{nn})^{T}.
\end{equation}
Equivalently, if
\begin{equation}
A=\sum_{k,l=1}^{n} a_{kl}\,|k\rangle\langle l|,
\end{equation}
then
\begin{equation}
\mathrm{vec}[A] = (I_{n\times n}\otimes A)\sum_{i=1}^{n} e_i\otimes e_i,
\end{equation}
where $e_i$ are the canonical basis vectors of $\mathbb{C}^n$. In other words, the \emph{vec}-operator transforms a matrix into a column vector by stacking its columns one after another.

Using standard identities for the Kronecker product, one obtains
\begin{align}
\mathrm{vec}[AB] &= (I_n\otimes A)\,\mathrm{vec}[B]
= (B^{T}\otimes I_n)\,\mathrm{vec}[A], \\
\mathrm{tr}(A^{\dagger}B) &= \mathrm{vec}[A]^{\dagger}\mathrm{vec}[B], \\
\mathrm{vec}[AXB] &= (B^{T}\otimes A)\,\mathrm{vec}[X],
\end{align}
for arbitrary matrices $A$, $B$, and $X$.

Before presenting the explicit vectorized form of the QFIM associated with the density matrix $\rho$, we briefly recall the definition of the quantum Fisher information. For a state $\rho_{\theta}$ depending on a single parameter $\theta$, the quantum Fisher information is given by
\begin{equation}
F(\rho_\theta)=\mathrm{Tr}\!\left(\rho_\theta L_\theta^2\right),
\end{equation}
where $L_\theta$ denotes the symmetric logarithmic derivative (SLD), defined implicitly by
\begin{equation}
2\,\partial_\theta \rho = L_\theta \rho + \rho L_\theta.
\end{equation}
When several parameters $\{\theta_i\}=\{\theta_1,\theta_2,\ldots,\theta_n\}$ are estimated simultaneously, the relevant object is the QFIM,
\begin{equation}
F_{ij}=\frac{1}{2}\mathrm{Tr}\!\left[\left(L_{\theta_i}L_{\theta_j}+L_{\theta_j}L_{\theta_i}\right)\rho\right].
\end{equation}
Its explicit construction requires solving the SLD equations above. Several equivalent expressions for the QFIM are available in the literature.

Using the spectral decomposition
\begin{equation}
\rho=\sum_k p_k |k\rangle\langle k|,
\end{equation}
the QFIM can be written as
\begin{equation}
F_{ij}
=
2\sum_{p_k+p_l>0}
\frac{
\langle k|\partial_{\theta_i}\rho|l\rangle
\langle l|\partial_{\theta_j}\rho|k\rangle
}{p_k+p_l},
\end{equation}
while the SLDs take the form
\begin{equation}
L_{\theta_i}
=
2\sum_{p_k+p_l>0}
\frac{\langle k|\partial_{\theta_i}\rho|l\rangle}{p_k+p_l}
|k\rangle\langle l|.
\end{equation}
Another useful representation expresses the QFIM through the exponential of the density matrix:
\begin{equation}
F_{ij}=2\int_0^\infty
\mathrm{Tr}\!\left(
e^{-\rho t}\,\partial_{\theta_i}\rho\,
e^{-\rho t}\,\partial_{\theta_j}\rho
\right)\,dt.
\end{equation}

More recently, an explicit formula based on the vectorization of $\rho$ was introduced. This approach is analytically convenient for arbitrary systems and avoids both diagonalization, as in the spectral form above, and integral representations, as in the previous equation. It relies on the inverse of the matrix
\begin{equation}
\Lambda=\rho^T\otimes I + I\otimes \rho.
\end{equation}

Using the identities above, one finds that the QFIM can be rewritten as
\begin{equation}
F_{ij}=2\,\mathrm{vec}[\partial_i \rho]^{T}\Lambda^{-1}\mathrm{vec}[\partial_j \rho],
\end{equation}
and the SLDs are given by
\begin{equation}
\mathrm{vec}[L_{\theta_i}]
=
2\,\Lambda^{-1}\mathrm{vec}[\partial_i\rho].
\end{equation}

In single-parameter estimation, the scalar Cramér--Rao bound,
\begin{equation}
\mathrm{Var}(\theta)\geq F^{-1},
\end{equation}
is in general saturable. The optimal measurement is then associated with the projectors onto the eigenvectors of the SLD operator $L_\theta$. In contrast, in the multiparameter case the matrix inequality
\begin{equation}
\mathrm{Cov}(\hat{\boldsymbol{\theta}})\geq F^{-1}
\end{equation}
is not always saturable, since the optimal measurements for different parameters may be incompatible. It is therefore important to determine the conditions under which the bound can be reached. If the SLDs commute, i.e.
\begin{equation}
[L_{\theta_i},L_{\theta_j}]=0,
\end{equation}
then a common eigenbasis exists and simultaneous measurements can saturate the bound. This commutativity condition is sufficient but not necessary. When the SLDs do not commute, a weaker sufficient condition for saturability is
\begin{equation}
\mathrm{Tr}\!\left[\rho\,[L_{\theta_i},L_{\theta_j}]\right]=0,
\end{equation}
which guarantees the attainability of the Cramér--Rao bound in multiparameter estimation.

Here, we consider the optimal estimation of the temperature $T$ of a bosonic thermal bath by means of a quantum probe interacting with the bath, which plays the role of the environment. The state of the probe is described by a density operator $\rho$, and due to the interaction with the surroundings it acquires an explicit dependence on the bath temperature, $\rho \to \rho_T$. In this setting, $T$ is not the temperature of the probe itself, but rather a parameter encoded in its quantum state. This is in contrast with classical thermometry, where the probe is typically allowed to thermalize with the sample until equilibrium is reached, so that the temperature inferred from the probe coincides with that of both systems. In the quantum case, the larger family of accessible states, together with the strong sensitivity of quantum states to decoherence, can in principle make thermometric protocols more accurate than their classical counterparts.\par

To estimate the parameter $T$, one performs $M$ independent measurements of an observable $X$ on identically prepared probes. One convenient spectral representation of the QFI is
\begin{equation}
H(T)=2\sum_{m,n}
\frac{\left|\langle \psi_m|\partial_T \rho_T|\psi_n\rangle\right|^2}{\lambda_m+\lambda_n},
\end{equation}
where $\{|\psi_n\rangle\}$ and $\{\lambda_n\}$ are respectively the eigenvectors and eigenvalues of the temperature-dependent state $\rho_T$. The QFI can also be expressed in terms of the symmetric logarithmic derivative (SLD),
\begin{equation}
H(T)=\mathrm{Tr}\big[\rho_T L_T^2\big],
\end{equation}
with
\begin{equation}
L_T
=
2\sum_{m,n}
\frac{\langle \psi_m|\partial_T \rho_T|\psi_n\rangle}{\lambda_m+\lambda_n}
|\psi_m\rangle\langle \psi_n|.
\end{equation}
The SLD is particularly useful because the optimal measurement is given by its spectral decomposition.
To characterize the precision in estimating a parameter regardless of its actual magnitude, one can define the signal-to-noise ratio (SNR) as
\[
QSNR=\frac{T^2}{\mathrm{Var}(\hat T)},
\]
where larger values of \(R_T\) correspond to more accurate estimators. By employing the quantum Cramér--Rao bound, one obtains
\[
R_T \leq Q_T \equiv T^2 H(T),
\]
with \(H(T)\) denoting the quantum Fisher information associated with the parameter \(T\). The quantity \(Q_T\), commonly called the quantum signal-to-noise ratio (QSNR), provides a measure of the ultimate precision allowed by quantum mechanics: higher values of \(Q_T\) indicate that the parameter \(T\) can be estimated more efficiently.
A further figure of merit closely related to the QFI is the quantum signal-to-noise ratio (QSNR),
\begin{equation}
QSNR=T^2 H(T).
\end{equation}
This quantity compares the square of the parameter to the variance of its best possible estimate. A larger value of $R(T)$ corresponds to a smaller relative error and, consequently, to a more precise thermometric protocol.

This relation clearly indicates that increasing the QSNR leads to a reduction of the relative uncertainty, thereby enhancing the precision of temperature estimation. The optimization over the full set of possible quantum measurements can in principle be carried out. In what follows, the QFI and QSNR will serve as the main tools for assessing the performance of the temperature estimation scheme. In the following, we apply the formalism introduced above to the problem of estimating the temperature $T$ of a structured sample characterized by an Ohmic-type spectral density. The proposed thermometric protocol relies on a qubit that interacts with the sample for a finite time, after which a measurement is performed to extract information about the temperature. In particular, we examine whether a reliable quantum thermometry scheme can be realized within an exactly solvable dephasing model for the probe qubit. To this end, we determine the interaction time that maximizes the QFI and evaluate the associated QSNR. As shown below, the numerical results are favorable across all the cases considered, while analytic expressions can also be obtained in some regimes, namely in the super-Ohmic case, in the low-temperature Ohmic regime, and in the high-temperature limit for all spectral densities.

\begin{figure}[t] 
        \includegraphics[width=10.0cm]{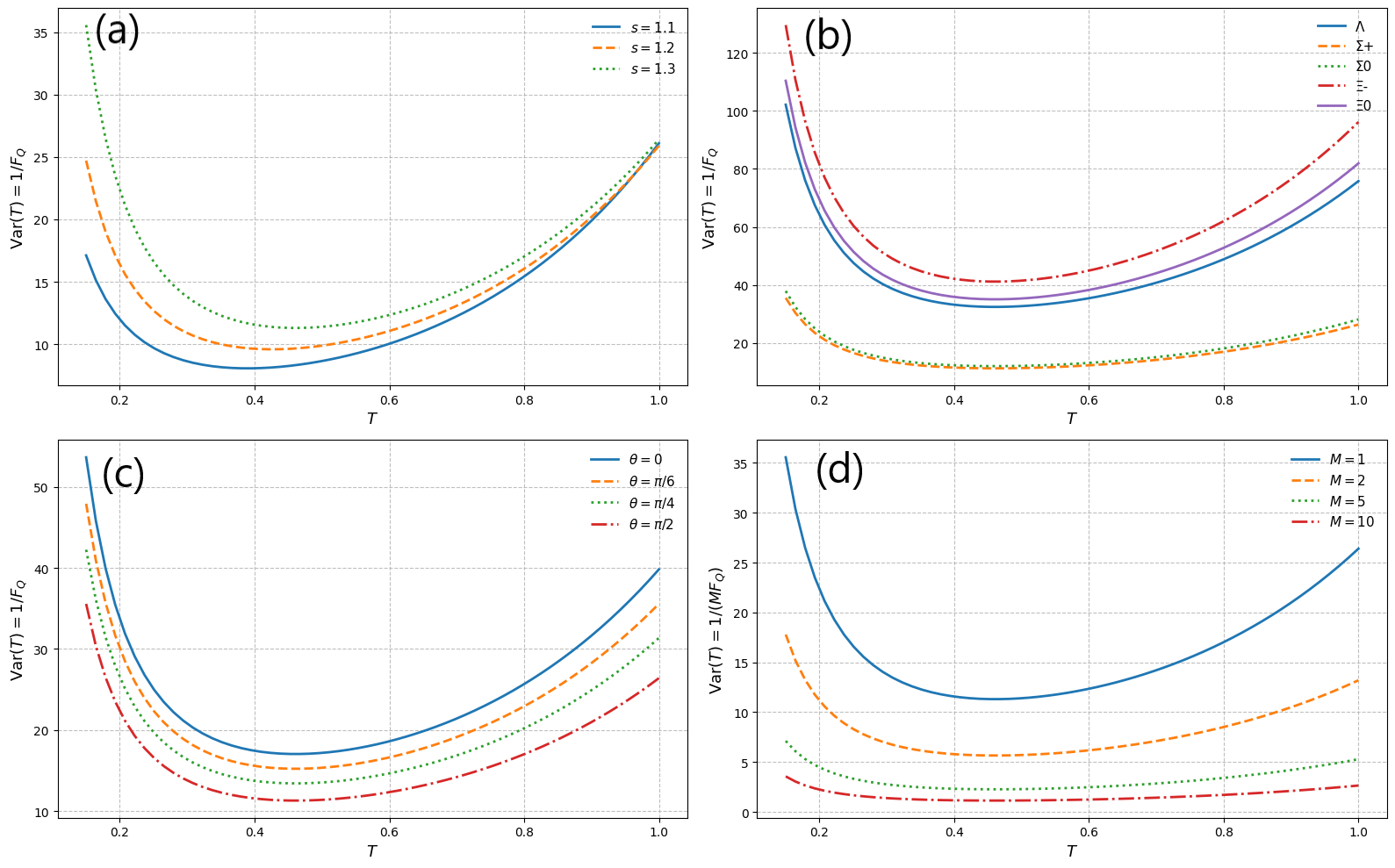}
      \caption{The minimal bound on the variances of the individual estimators of the parameter  \(T\) for different ohmicity $s$, hyperons $Y$, diffusion angle $\theta$ and the number of measurements M.}
    \label{fig3}
\end{figure}

\section{Quantum thermometry in ohmic reservoir}

The purpose of this section is to determine the optimal parameters allowing an efficient estimation of the temperature while minimizing the fluctuations associated with the measurement process. To this end, we first investigate the minimum variance of $T$ as a function of temperature for different collision parameters and for the various outgoing particle species, namely hyperons, in order to reduce measurement errors. Next, we analyze the QFI and QSNR for different Ohmicity parameters in order to examine their influence on temperature estimation. We also study the effects induced by local and mutual dephasing on the probe and perform a comparison of the temperature-estimation efficiency for the different hyperons. Finally, we conclude by evaluating the optimal interaction time $t_{\mathrm{opt}}$ and the optimal temperature $T_{\mathrm{opt}}$, which arise from the presence of finite peaks for different Ohmicity regimes.

\subsection{Quantum measurement error}

The dynamics induced by the functions \(\Gamma_s(\tau,T)\) and \(f(\tau)\) corresponds to a collective dephasing channel with a dynamical phase given by

\begin{equation}
\rho_{\mathrm{CB}}(\tau,T)=V(\tau,T)\circ R(\tau)\circ \rho,
\end{equation}

The dynamical density matrix have the following expression:

\[
\rho =
\begin{pmatrix}
a & 0 & 0 & x \\
0 & b & z & 0 \\
0 & z & b & 0 \\
x & 0 & 0 & d
\end{pmatrix}
\]

To evaluate the quantum Fisher information matrix, we have first to compute the matrix $\Lambda$ :
\[
\Lambda =
\begin{pmatrix}
\Lambda_{11} & 0_{4\times4} & 0_{4\times4} & \Lambda_{14} \\
0_{4\times4} & \Lambda_{22} & \Lambda_{23} & 0_{4\times4} \\
0_{4\times4} & \Lambda_{32} & \Lambda_{33} & 0_{4\times4} \\
\Lambda_{41} & 0_{4\times4} & 0_{4\times4} & \Lambda_{44}
\end{pmatrix},
\tag{19}
\]

with $\Lambda_{ij}$ $(i,j = 1,2,3,4)$ are the $4 \times 4$ matrix given by
\[
\Lambda_{11} =
\begin{pmatrix}
2a & 0 & 0 & x\\
0 & a+b & z & 0\\
0 & z & a+b & 0\\
x & 0 & 0 & a+d
\end{pmatrix},\quad 
\Lambda_{14}
\begin{pmatrix}
x & 0 & 0 & 0 \\
0 & x & 0 & 0 \\
0 & 0 & x & 0 \\
0 & 0 & 0 & x
\end{pmatrix}.
\tag{20}
\]

and
\[
\Lambda_{22} =
\begin{pmatrix}
a+b & 0 & 0 & x\\
0 & 2b & z & 0\\
0 & z & 2b & 0\\
x & 0 & 0 & b+d
\end{pmatrix},
\quad 
\Lambda_{23} =
\begin{pmatrix}
z & 0 & 0 & 0 \\
0 & z & 0 & 0 \\
0 & 0 & z & 0 \\
0 & 0 & 0 & z
\end{pmatrix},
\tag{21}
\]
\[
\Lambda_{44}=
\begin{pmatrix}
a+d & 0 & 0 & x\\
0 & b+d & z & 0\\
0 & z & b+d & 0\\
x & 0 & 0 & 2d
\end{pmatrix}.
\]
with \[
\Lambda_{14}=\Lambda_{41} \quad \quad \quad \Lambda_{23}=\Lambda_{32}\]
The inverse of the matrix $\Lambda$ Eq.(19) is given by
\[
\Lambda^{-1} =
\begin{pmatrix}
(\Lambda^{-1})_{11} & 0_{4\times4} & 0_{4\times4} & (\Lambda^{-1})_{14} \\
0_{4\times4} & (\Lambda^{-1})_{22} & (\Lambda^{-1})_{23} & 0_{4\times4} \\
0_{4\times4} & (\Lambda^{-1})_{32} & (\Lambda^{-1})_{33} & 0_{4\times4} \\
(\Lambda^{-1})_{41} & 0_{4\times4} & 0_{4\times4} & (\Lambda^{-1})_{44}
\end{pmatrix},
\tag{22}
\]

with
\[
(\Lambda^{-1})_{11} =
\begin{pmatrix}
\eta' & 0 & 0 & \theta'\\
0 & \dfrac{\Delta}{\lambda} & \Delta\omega & 0 \\
0 & \Delta\omega & \Delta\lambda & 0 \\
\theta ' & 0 & 0 & \zeta'
\end{pmatrix},
(\Lambda^{-1})_{22} =
\begin{pmatrix}
\dfrac{\lambda}{\Delta} & 0 & 0 & \chi \\
0 & \sigma'& \tau & 0 \\
0 & \tau & \sigma' & 0 \\
\chi & 0 & 0 & \dfrac{\pi}{\Delta}
\end{pmatrix},
\]

and
\[
(\Lambda^{-1})_{23} =
\begin{pmatrix}
\dfrac{\omega}{\Delta} & 0 & 0 & \phi \\
0 & \tau & \upsilon & 0 \\
0 & \upsilon & \tau & 0 \\
\phi & 0 & 0 & \dfrac{\psi}{\Delta}
\end{pmatrix},
\quad 
(\Lambda^{-1})_{14} =
\begin{pmatrix}
\theta'& 0 & 0 & \kappa'\\
0 & \Delta\chi & \Delta\phi & 0 \\
0 & \Delta\phi & \Delta\chi & 0 \\
\kappa' & 0 & 0 & \iota'
\end{pmatrix}.
\]

\[
(\Lambda^{-1})_{44}=
\begin{pmatrix}
2\delta'+\kappa' & 0 & 0 & \iota' \\
0 & \Delta\pi & \Delta\psi & 0 \\
0 & \Delta\psi & \Delta\pi & 0 \\
\iota' & 0 & 0 &\zeta'
\end{pmatrix}.
\]

with \[
(\Lambda^{-1})_{14}=(\Lambda^{-1})_{41} \quad \quad \quad (\Lambda^{-1})_{23}=(\Lambda^{-1})_{32}\]

\begin{align}
\operatorname{vec}[\partial_{T}\rho]
&=
\begin{aligned}[t]
\bigl(&\partial_{T}a,0,0,\partial_{T}x,0,\partial_{T}b,\partial_{T}z,0,\\
     &0,\partial_{T}z,\partial_{T}b,0,\partial_{T}x,0,0,\partial_{T}d\bigr)^{T}
\end{aligned}
\label{eq:vec_dT}
\end{align}

The QFI matrix can be determined as
\begin{equation}
F_{TT}
=
\begin{aligned}
&2\,\operatorname{vec}[\partial_{T}\rho]^{T}\,\Lambda^{-1}\times\,\operatorname{vec}[\partial_{T}\rho]
\end{aligned}
\end{equation}

Our aim is to perform an individual estimation of the variance in T in order to determine the optimal internal parameters that will minimize the estimation error by selecting appropriate optimal parameters.
In the individual estimation scenario, each parameter is inferred separately under the assumption that the parameters are statistically independent. In this case, estimating one parameter precisely does not influence the estimation accuracy of the others. This situation arises only when the off-diagonal elements of the Fisher information matrix vanish, namely $F_{ij}=0$ for $i\neq j$. Under this condition, the corresponding variance bounds reduce to
\begin{equation}
\mathrm{Var}(T)_{\mathrm{Ind}} \geq 1/F_{TT}.
\tag{49}
\end{equation}

\begin{figure}[t] 
        \includegraphics[width=10cm]{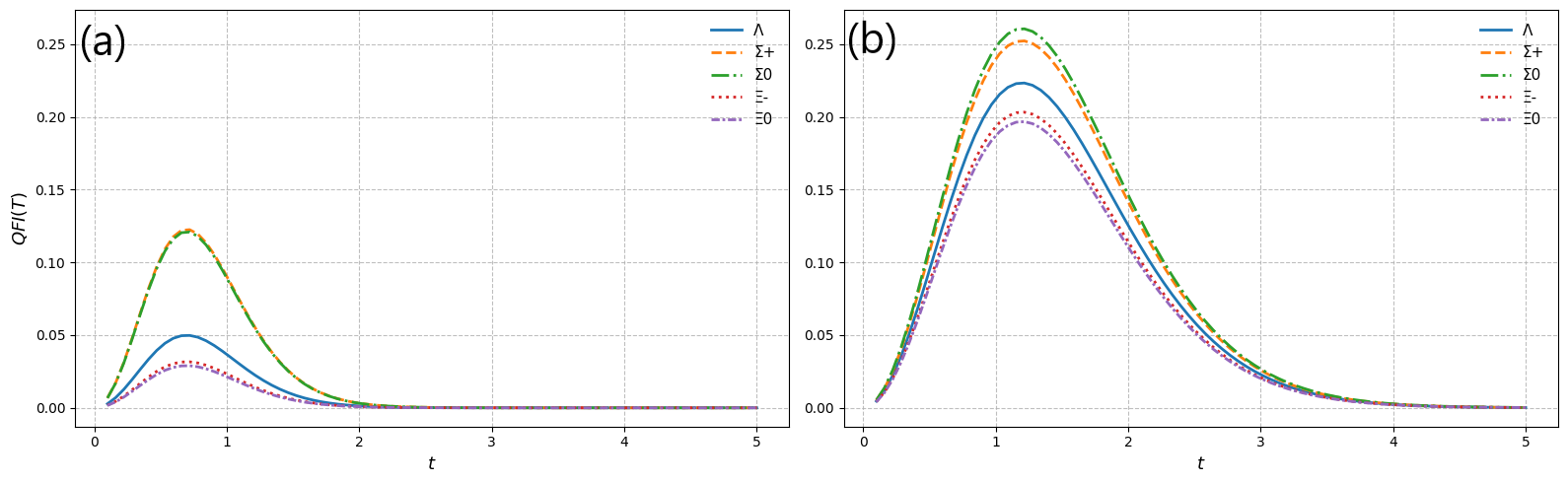}
      \caption{In the upper panel, we plot the contour lines of the ratio \(R\) associated with the \(\Sigma\) hyperon for different Ohmicities, namely \(s=1\) and \(s=3\). In the lower panel, we present the QFI as a function of time \(t\) for different hyperons in order to identify the optimal hyperon exhibiting the highest thermal sensitivity.}
    \label{fig4}
\end{figure}

The minimal variance of the temperature for individual estimation is then:
\begin{equation}
\mathrm{Var}(T)_{\mathrm{Ind}} = 1/F_{TT}.
\tag{49}
\end{equation}

\begin{figure*}[t] 
        \includegraphics[width=5.8cm]{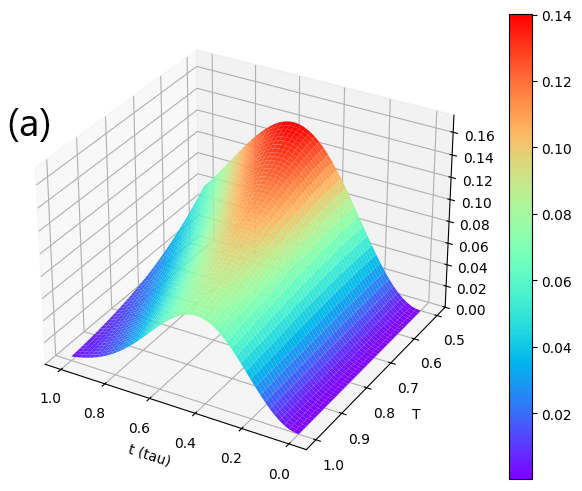}
        \includegraphics[width=5.8cm]{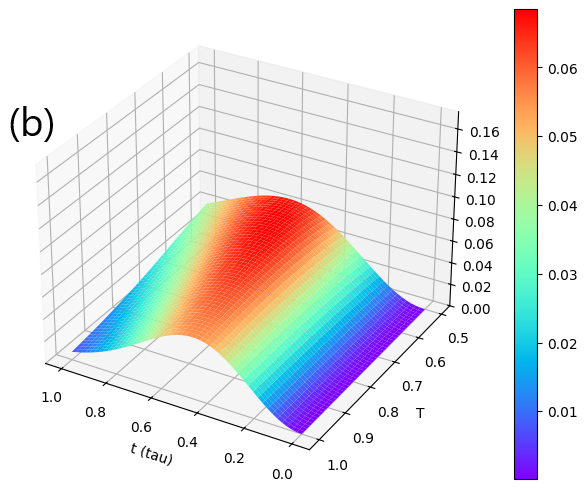}
        \includegraphics[width=5.8cm]{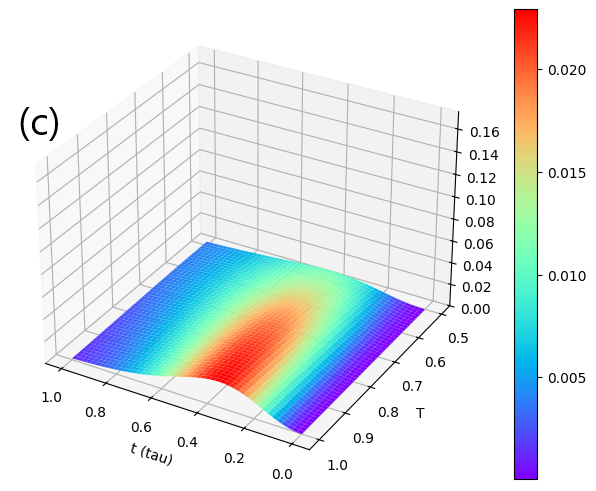}
      \caption{We show the QFI $H(T, t)$  as a function of both the reservoir temperature $T$ and the interaction time $t$ for three illustrative types of structured environments: (a) ($s = 1$), (b)  ($s = 1.5$), and (c)  ($s = 2.5$).}
    \label{fig5}
\end{figure*}

It is simple to verify that

\[
\begin{aligned}
F_{TT} ={}&
\frac{1}{(a+d)(ad-x^2)}
\Big[
(ad+d^2-x^2)(\partial_T a)^2
\\[4pt]
&\quad
+(a^2+ad-x^2)(\partial_T d)^2
+(a^2+3ad)(\partial_T x)^2
\\[4pt]
&\quad
+2x^2 (\partial_T a)(\partial_T d)
-4dx (\partial_T a)(\partial_T x)
\\[4pt]
&\quad
-4ax (\partial_T x)(\partial_T d)
\Big]
\\[8pt]
&+
\frac{2}{b^2-z^2}
\Big[
b\big((\partial_T b)^2+(\partial_T z)^2\big)
-2z (\partial_T b)(\partial_T z)
\Big].
\end{aligned}
\]

The analysis of the individual variance in the estimation of $T$ makes it possible to identify an optimal choice for different ohmicities, deviation angles $\theta$, and the hyperon Y  with different decay parameters ($\alpha_\psi$,$\beta_\psi$), thereby minimizing the estimation errors in our thermometric study by choosing the optimal parameters. The Fig.\ref{fig3}(a) illustrates the behavior of the minimal variances obtained within the framework of separate temperature estimations, denoted by $T$. At low temperatures, the variance of $T$ reaches its minimum when the parameter lies within the interval $I_c \in [0.4,\,0.6]$.In contrast, at high temperatures, the variance increases drastically, leading to a significantly less accurate estimation. This degradation is primarily due to strong thermal fluctuations, which hinder precise measurements and drive the system away from the confidence interval. The increase in ohmicicity shifts the optimal temperature from 0.4 to 0.5 within the confidence interval. This shift arises from the displacement of the estimation peaks toward denser thermal regimes, which compensates for the hindrance caused by ohmicicity. At high temperatures, this effect eventually converges, losing its dependence on the ohmic parameter $  s  $, and exhibits a monolithic behavior that tends to increase the variance.
Similarly, the scattering angle and the production angle in Fig.\ref{fig3}(b-c) shift the confidence interval from 0.4 to 0.5  $\theta = 0$, with a minimal variance for $  \theta = \pi/2  $ and small values of the decay parametres $(\alpha_\psi,\beta_\psi )  $. In this regime, the increase leads to a decrease in precision due to the rise in var(T). These results allow us to hypothesize that hyperons with negative $\alpha$ and $\beta$, such as $  \Sigma  $ and $  \Sigma^0 $, slow down the decoherence process and exhibit better thermal adherence compared to other particles. To verify this assumption, we plotted the QFI as a function of time for the different hyperons in order to determine which particle provides the most efficient temperature estimation for mutual and local bath. As clearly shown in Fig.\ref{fig4}(a-b), the $\Sigma$ and $\Sigma^{0}$ hyperons exhibit the largest QFI values, indicating a higher thermal sensitivity. We have examined the minimum variance for all the internal parameters while keeping the number of measurements fixed. In the following, we vary the number of measurements in order to investigate how increasing the number of estimations affects the uncertainty. In Fig \ref{fig3}(d) shows the optimal relative error, as a function of the temperature $T$, for 1, 2, 5, and 10 measurements. It is clear that, in the low-temperature regime, the optimal relative error depends significantly on $T$, indicating that this strategy is particularly well suited to the design of quantum sensors operating at low temperatures. In particular, increasing the number of measurements leads to a drastic reduction in the relative error, which implies higher precision and a smaller variance. By contrast, in the high-temperature regime, the relative error increases substantially due to thermal agitation, which makes both the measurement and the estimation of temperature more challenging.

\begin{figure*}[t] 
          \includegraphics[width=5.8cm]{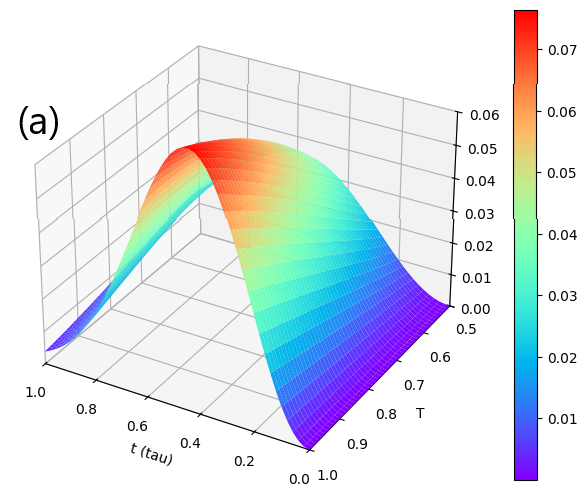}
        \includegraphics[width=5.8cm]{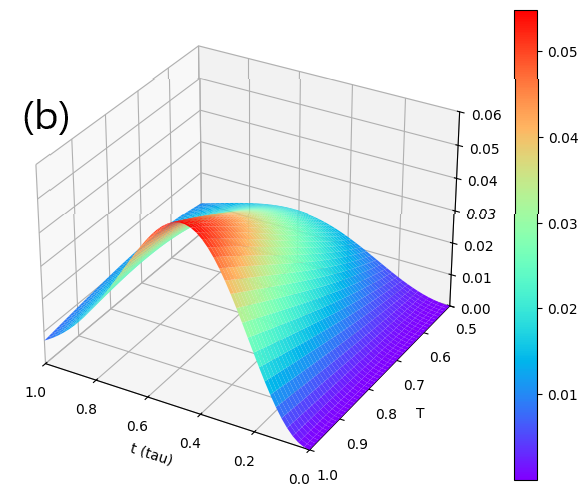}
        \includegraphics[width=5.8cm]{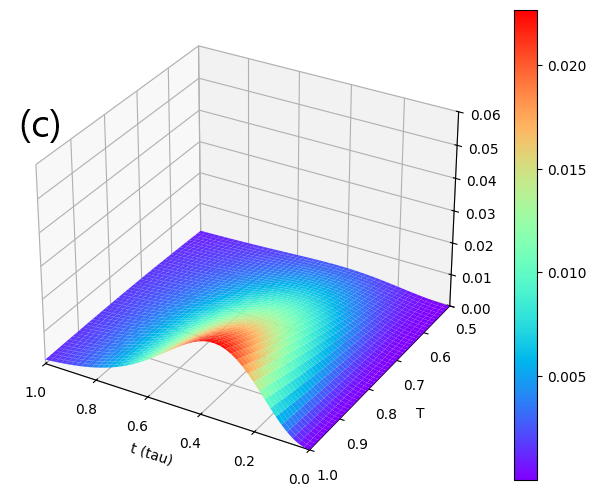}
      \caption{Quantum signal noise QSNR as a function of both the reservoir temperature $T$ and the interaction time $t$ for three illustrative types of structured environments: (a) ($s = 1$), (b)  ($s = 1.5$), and (c)  ($s = 2.5$).}
    \label{fig6}
\end{figure*}

\subsection{Thermal quantum estimation}

In the following, we investigate quantum thermometry using two-qubits probe coupled to environments. Our approach employs a probe that interacts with the sample for a controlled duration, after which a suitable measurement is performed on the qubit to infer the temperature. We focus on different representative values of the Ohmicity parameter $s$.We apply this formalism to estimate the temperature \( T \) of a structured thermal reservoir characterized by an Ohmic-type spectral density.\par

The Fig.\ref{fig5}(a-c) display three-dimensional plots of the QFI, as a function of both temperature $T$ and interaction time $t$, for  Ohmic ($s=1$) environments, the QFI exhibits a clear maximum at a finite time particularly for lower temperature $T$,  in contrast to the super-Ohmic case ($s=1.5,2$) where the QFI reaches its highest values at elevated temperatures, while at low temperatures it saturates without displaying a distinct peak at a specific interaction time, it is clear that the Ohmicity parameter $s$ tends to flatten the peak and deteriorate the thermal estimation performance of the probe. In order to better assess the bounds of the estimation and the influence of internal parameters, it is essential to analyze the behavior of the quantum signal-to-noise ratio (QSNR).\par

Where it serves as a key figure of merit in quantum metrology, as it directly compares the strength of the encoded signal to the underlying quantum fluctuations. A higher QSNR implies that the parameter \( T \) can be estimated with greater precision and efficiency.\par

The QSNR shown in Fig.~\ref{fig6}(a-c) is computed for three representative values of the Ohmicity parameter discussed earlier. In the low-temperature regime, the QSNR tends to vanish, reflecting the poor performance of the estimation process in this limit. As the temperature increases, the QSNR rises significantly. In the intermediate temperature region, its behavior depends explicitly on the value of $s$, leading to different trends for different spectral profiles. At sufficiently high temperatures, however, the influence of the environmental structure becomes negligible, and the QSNR approaches a common saturation value, independent of the spectral density. We shall next carry out a comparative study to examine the effect of mutual and local dephasing on temperature estimation.

\begin{figure}[t] 
        \includegraphics[width=10cm]{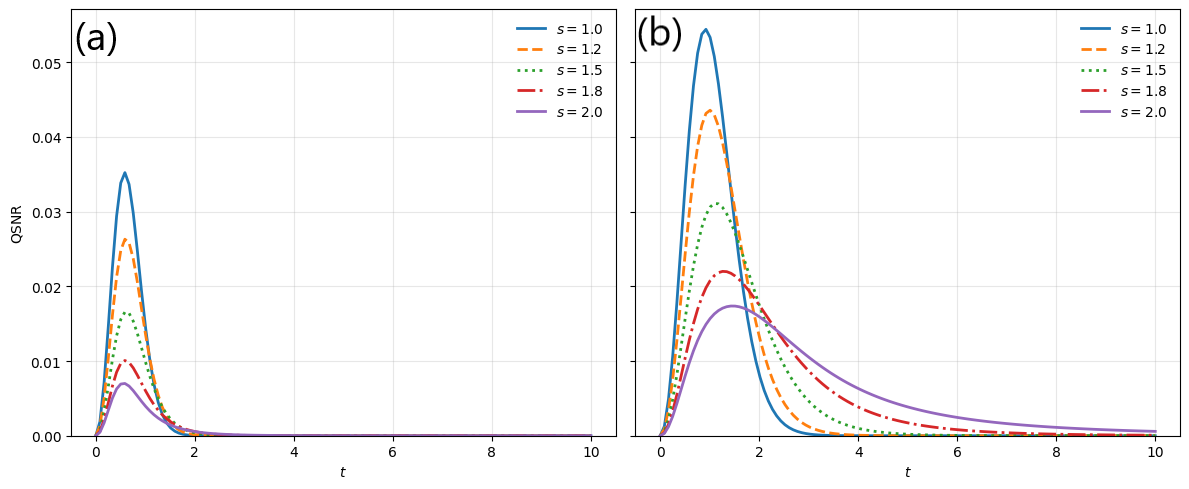}
        \includegraphics[width=9.8cm]{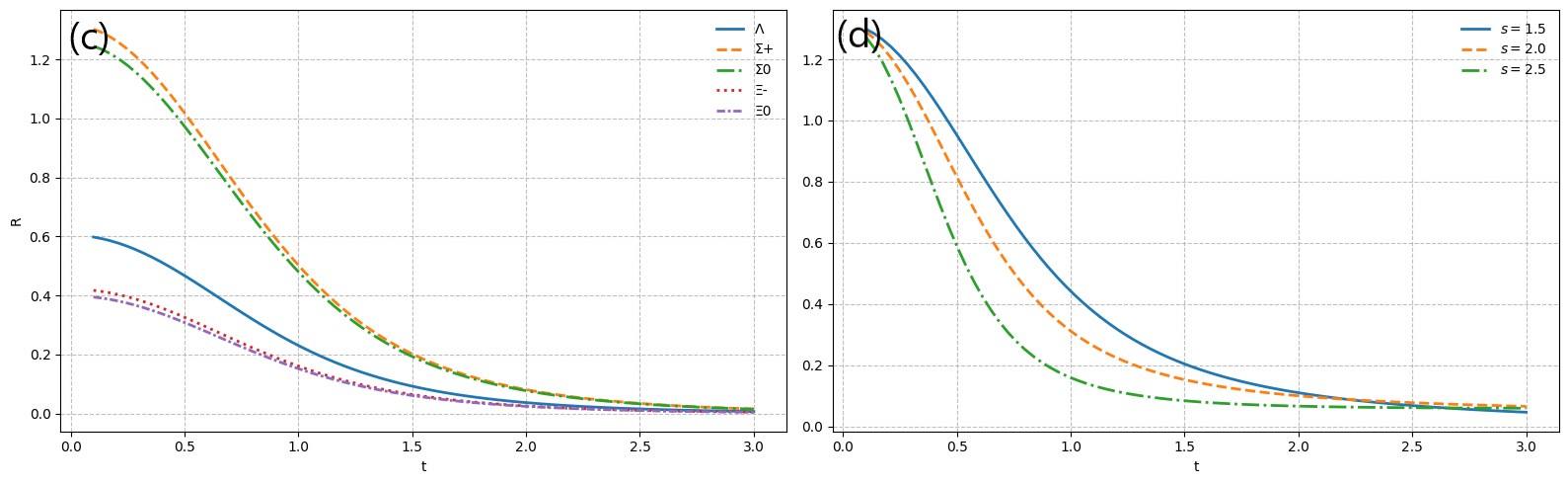}
      \caption{In the upper panel, we present the QSNR for the case of a dynamics induced by mutual and local dephasing for different ohmic regimes as a function of evolution time t . We then define the ratio \(R\), given by the QSNR under mutual dephasing divided by the QSNR under local dephasing, for different hyperons. We therefore fix the hyperon to \(\Sigma\) and plot \(R\) as a function of time t for different Ohmicities s.}
    \label{fig7}
\end{figure}

To establish a reference framework, we fixed the hyperon to $\Sigma$ and plotted the QSNR for two probes subjected to both mutual and local dephasing mechanisms in Fig.\ref{fig7}(a-b). It is evident that the overall behavior, as well as the influence of Ohmicity, remains qualitatively similar in both scenarios. However, the main difference lies in the magnitude of the response, which is noticeably weaker in the mutual-dephasing case. This reduction originates from the interference effects between the two probes, which enhance the decoherence process and consequently lead to a less accurate temperature estimation compared with the local-dephasing scenario.Where for very short interaction times, mutual dephasing appears to be more advantageous, yielding a more favorable estimation performance and reaching equilibrium more rapidly than local dephasing. By contrast, local dephasing becomes more efficient at intermediate times, although it relaxes more slowly toward equilibrium.\par

To verify this point let's introduce the factor \(R\), defined as the ratio between the mutual QSNR and the local QSNR in Fig.\ref{fig7}(c-d). For short evolution times ($\tau \ll 1$), the highest precision in temperature estimation is achieved by using  $\Sigma^+$, $\Sigma^0$ as a quantum probe coupled to a common thermal bath. In contrast, for longer times, superior estimation accuracy is obtained when the probe evolves under the influence of independent local baths. This can be explained by the presence of mutual decoherence, which occurs faster than local decoherence and leads to more accurate thermal estimation along with enhanced stability.
Particular for the weak-ohmic regime that governs the timescale required for the probe to reach equilibrium and provides a longer temporal window for thermal sensitivity. The observed advantage of the common bath configuration for $\Sigma^0$ can be attributed to the constructive role of bath-induced correlations, which enhance the temperature sensitivity of the probe state due to the presence of negative diffusion coefficients, $\alpha_{\psi}$ and $\beta_{\psi}$, which effectively slow down the decoherence process and thereby enhance the feasibility of the temperature estimation.

In contrast, for particles characterized by positive diffusion coefficients  $\Xi^-$ and $\Xi^0$, the same collective interaction generates correlations that are either weakly dependent on temperature or detrimental to the estimation process, resulting in no improvement over the local bath scenario because the mutual dephasing degrades the estimation process. This highlights the crucial interplay between the system parameters and the structure of environment-induced correlations in quantum thermometry, with a more efficient estimation in the case of which is also less sensitive to the mutual noise present in the reservoir.\par

In this subsection, we have analyzed the estimation of the temperature in the Ohmic reservoir and clearly observed the existence of finite temporal and thermal peaks. This, in turn, supports the existence of an optimal parameter regime for maximizing thermal sensitivity.

The presence of a maximum in the QFI indicates that optimal temperature estimation can be achieved at a finite optimal interaction time $t_{\rm opt}$, before the probe reaches thermal equilibrium with the environment.

\subsection{Optimal bounds and performance}

 For clarity, we restrict our discussion to the $\Sigma\bar{\Sigma}$ channel. This choice is motivated by the fact that it captures the essential thermal behavior observed in other hyperon systems. The presence of an optimal interaction time and an optimal temperature is a central characteristic of quantum thermometry. For very short interaction times, the probe does not have enough time to encode the thermal properties of the environment, and the resulting estimation precision remains low. As the interaction time increases, both the quantum Fisher information and the quantum signal-to-noise ratio increase, until they attain a maximum at a finite optimal time, \(t_{\mathrm{opt}}\). A comparable trend appears with temperature: the sensitivity is highest within a finite temperature range, whereas at sufficiently high temperatures, the system approaches thermal equilibrium, the sensitivity saturates, and the response becomes essentially independent of the geometry and the details of the environment. The corresponding optimal temperature is denoted by \(T_{\mathrm{opt}}\).\par

\begin{figure}[t]
        \includegraphics[width=9.3cm]{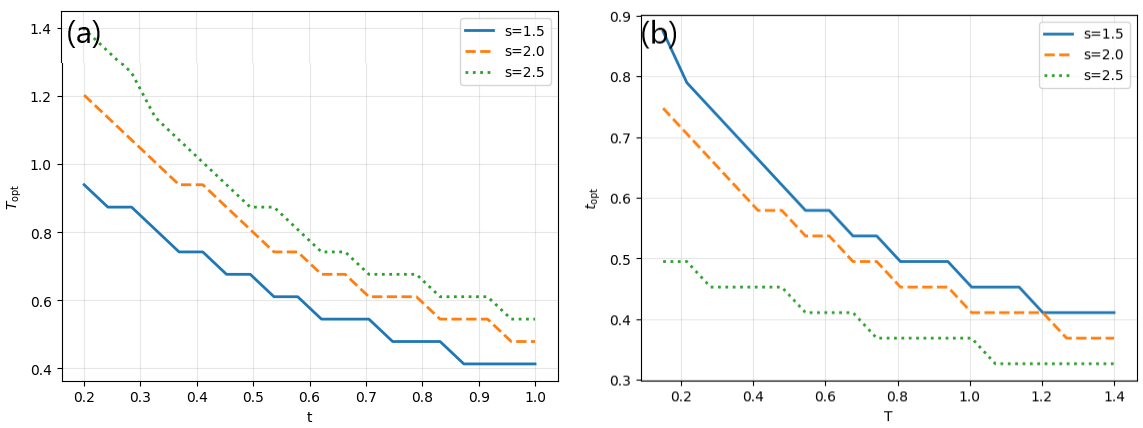}
         \includegraphics[width=9.3cm]{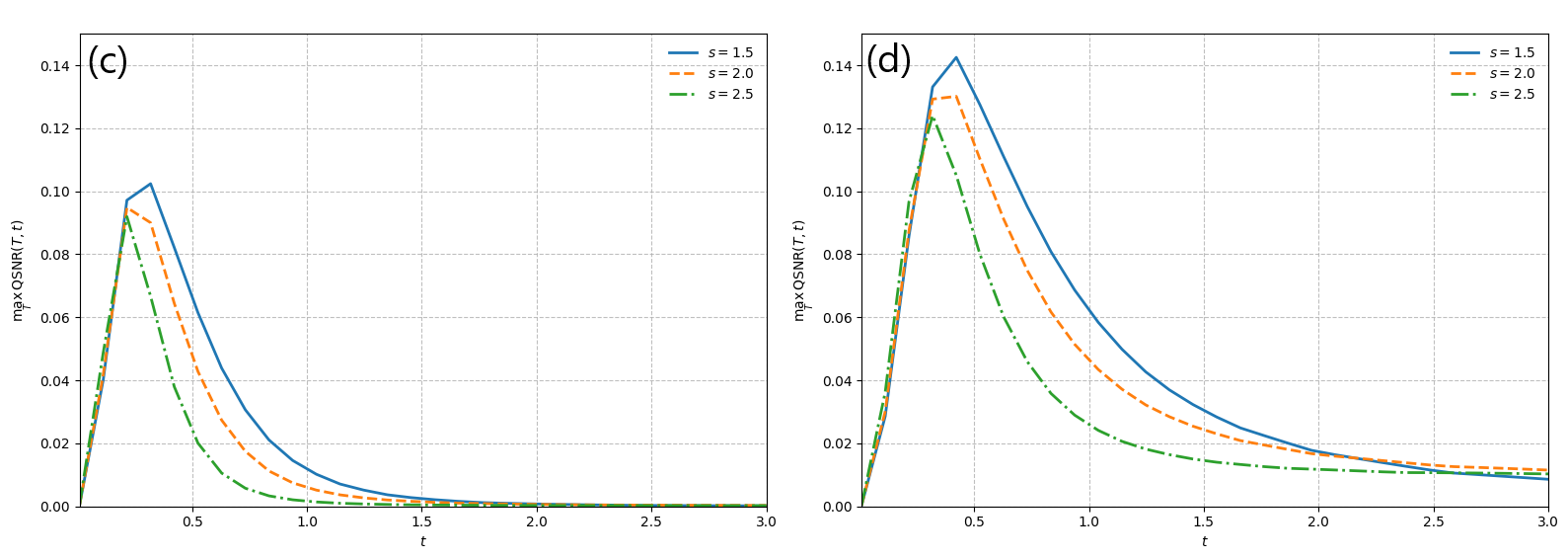}
      \caption{We present the optimal interaction times \(t_{\mathrm{opt}}\) as a function of the mutuel reservoir at temperature \(T\). In particular, we determine the temperature \(T_{\mathrm{opt}}\) at which the QFI attains its maximum and display it as a function of the interaction time \(t\). We also display, in the lower panel, the QSNR for both the common and local dephasing environments at the optimal temperature $T_{\mathrm{opt}}$ as a function of the evolution time $t$ for different values of the Ohmicity parameter $s$.}
    \label{fig8}
\end{figure}

We clearly observe in Fig.\ref{fig8}(a) that for low temperatures, the optimal probing time is significantly longer, whereas it decreases steadily as temperature increases. This trend is physically intuitive, as the probe extracts information about the environment through decoherence. At low temperatures, decoherence is slower, requiring more time for the environment to imprint its temperature information onto the qubit. 

A monotonic decrease from the initial values is observed. Notably, an increase in the ohmicity parameter \( s \) leads to a faster decay rate, thereby destroying the quantum coherence of the probe more rapidly, the weaker thermal fluctuations result in slower decoherence, necessitating longer interaction times for the temperature information to be effectively encoded into the qubit state through system-reservoir interactions. Conversely, at low temperatures the spectral properties of the environment play a decisive role in the decoherence dynamics, leading to a strong dependence of the QFI on the Ohmicity parameter $s$. At high temperatures, decoherence occurs rapidly due to stronger thermal fluctuations, making the detailed structure of the environment less relevant, where the QFI displays similar behavior across all three values of $s$ in the high-temperature regime.\par

The lower panels of Fig.~\ref{fig8}(b) display $T_{\mathrm{opt}}$, namely the temperature that maximizes the quantum Fisher information (QFI) for a given interaction time $t$. This optimal temperature decreases progressively as decoherence accumulates over time. As observed previously, increasing the Ohmicity parameter $s$ tends to flatten the QFI peaks and shift their maxima toward higher temperatures. Consequently, the super-Ohmic regime ($s=3.0$) requires significantly higher temperatures to achieve optimal estimation. This behavior stems from the stronger effective impedance imposed by the reservoir at high frequencies, which makes the probe less efficient under intense thermal conditions. Subsequently, we investigate the impact of the optimal parameters on the ultimate estimation bound for both the common and independent reservoirs.\par

The issue of the global estimability of the temperature is addressed in Fig.~\ref{fig8}(c-d), where we present the quantum signal-to-noise ratio (QSNR) evaluated at the optimal interaction time $t_{\rm opt}$. Specifically, we show the QSNR for different values of the Ohmicity parameter $s$, ranging from $1.5$ to $3$. It is clearly seen that at eraly times, the QSNR tends to zero, indicating that the efficiency of any estimation protocol is very poor in this regime. By contrast, it increases at higher times t. The dependence on the parameter $s$ is relevant only in the intermediate regime; near the equilibrium, we have a pronounced deterioration of the performance and the QSNR saturates to a universal value independent of the nature of the environmental spectral density.

\section{Quantum correlations}

In bipartite quantum systems, various types of quantum correlations reveal the non-classical features that play essential roles in quantum information processing. In the context of hyperon-antihyperon pairs, we will analyze several quantum features of the hyperon-antihyperon system by means of its spin density matrix, focusing on quantities such as quantum steering, entanglement of formation, geometric quantum discord, and quantum coherence.\par

Bell nonlocality represents the strongest form of quantum correlation and is revealed by the violation of Bell inequalities, demonstrating that quantum mechanics cannot be reproduced by any local hidden variable model. Quantum steering, which lies between Bell nonlocality and entanglement, describes the ability of one observer to remotely influence the state of a distant system through local measurements and is certified by the violation of the CJWR steering inequality, the entanglement is a central quantum resource and indicates the presence of non-separable states and serves as a key ingredient for quantum teleportation, cryptography, and computation,  it is quantified using concurrence, an entanglement monotone closely related to the entanglement of formation. Quantum discord, on the other hand, captures all quantum correlations by measuring the difference between two alternative expressions of quantum mutual information. These distinct resources characterize different aspects of quantum correlations and follow a well-established hierarchy: Bell nonlocality $\subset$ quantum steering $\subset$ entanglement $\subset$ quantum discord.\par

The spin density matrix of the hyperon-antihyperon system can be written as
\begin{equation}
\rho^{X}_{Y\bar{Y}}
=
\frac{1}{4}
\begin{pmatrix}
1+2a+t_3 & 0 & 0 & t_1-t_2 \\
0 & 1-t_3 & t_1+t_2 & 0 \\
0 & t_1+t_2 & 1-t_3 & 0 \\
t_1-t_2 & 0 & 0 & 1-2a+t_3
\end{pmatrix},
\end{equation}

The dynamics induced by the mutual dephasing channel is given by

\begin{equation}
\rho_{\mathrm{CB}}(\tau,T)=V(\tau,T)\circ R(\tau)\circ \rho,
\end{equation}

For a consistent and fair comparison, all correlation measures are normalized to the interval $[0,1]$. Furthermore, quantum coherence, which originates from the superposition principle of quantum states, constitutes another vital resource for quantum technologies. It is quantified here via the widely used $\ell_1$-norm of coherence, defined as
\[
C(\rho) = \sum_{i \neq j} |\rho_{i,j}|,
\]
The only nonvanishing off-diagonal elements are therefore
\[
\rho_{14},\ \rho_{41},\ \rho_{23},\ \rho_{32}.
\]

Since the density matrix is Hermitian, one has
\[
|\rho_{14}| = |\rho_{41}|,\qquad
|\rho_{23}| = |\rho_{32}|.
\]

Thus, the $\ell_1$-norm of coherence becomes
\[
C_{l_1}(\tau,T)
=
2|\rho_{14}|+2|\rho_{23}|.
\]

Substituting the dephased matrix elements yields
\[
|\rho_{14}|
=
\frac{
e^{-4\Gamma_s(\tau,T)}
|t_1-t_2|
}{4},
\qquad
|\rho_{23}|
=
\frac{
|t_1+t_2|
}{4}.
\]

Hence,
\[
C_{l_1}(\tau,T)
=
2\left(
\frac{
e^{-4\Gamma_s(\tau,T)}
|t_1-t_2|
}{4}
+
\frac{
|t_1+t_2|
}{4}
\right).
\]

Finally, one obtains
\[
C_{l_1}(\tau,T)
=
\frac{
e^{-4\Gamma_s(\tau,T)}
|t_1-t_2|
+
|t_1+t_2|
}{2}.
\]

Quantum discord provides a way to quantify the genuinely quantum part of correlations in a bipartite state. Its definition is based on the quantum mutual information of the density matrix $\rho_{AB}$,
\begin{equation}
I(A:B) \equiv S(\rho_A)+S(\rho_B)-S(\rho_{AB})
= S(\rho_A)-S(\rho_{A|B}),
\end{equation}
where
\begin{equation}
S(\rho)\equiv -\mathrm{Tr}\!\left(\rho \log_2 \rho\right)
\end{equation}
is the von Neumann entropy and
\begin{equation}
S(\rho_{A|B}) \equiv S(\rho_{AB})-S(\rho_B)
\end{equation}
is the conditional entropy. To evaluate discord, one introduces a local projective measurement on subsystem $B$, described by the set of one-dimensional projectors $\{\Pi_k\}$ satisfying $\sum_k \Pi_k = \mathbb{I}$. After obtaining the outcome $k$, the conditional state of subsystem $A$ is
\begin{equation}
\rho_{A|\Pi_k}
=
\frac{1}{p_k}
\mathrm{Tr}_B\!\left[
(\mathbb{I}\otimes \Pi_k)\,\rho_{AB}\,(\mathbb{I}\otimes \Pi_k)
\right],
\end{equation}
with probability
\begin{equation}
p_k = \mathrm{Tr}\!\left[
(\mathbb{I}\otimes \Pi_k)\,\rho_{AB}\,(\mathbb{I}\otimes \Pi_k)
\right].
\end{equation}
These conditional states define the ensemble $\{p_k,\rho_{A|\Pi_k}\}$. For a bipartite qubit system, one has $k=0,1$.

\begin{figure*}[t] 
        \includegraphics[width=18.8cm]{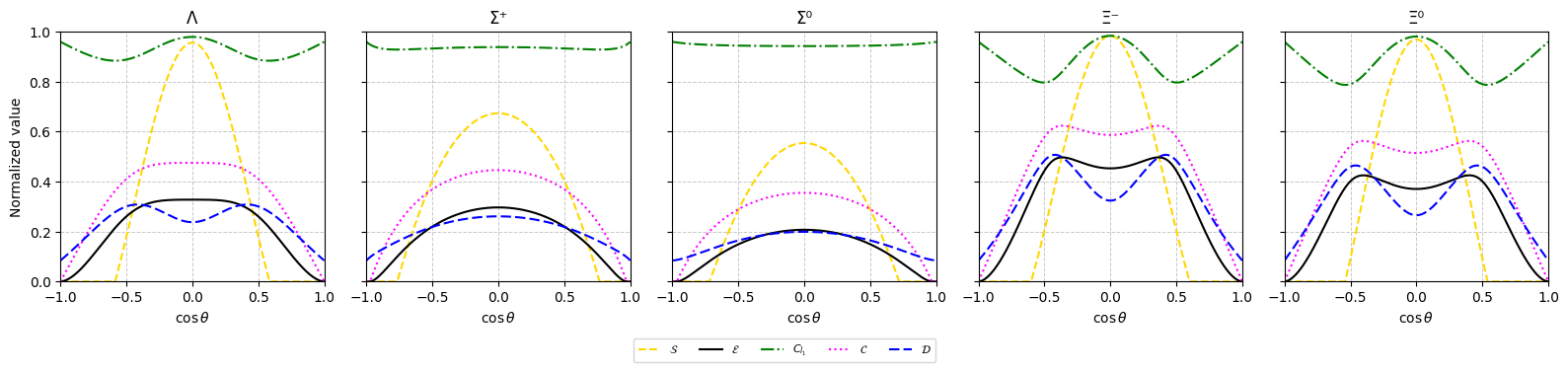}
      \caption{We display several quantum correlation measures as functions of \(\cos\vartheta\), where \(\vartheta\) denotes the scattering angle, for the process \(e^+e^- \to J/\psi \to Y\bar{Y}\). The five panels, labeled from (a) to (e), correspond respectively to \(\Lambda\), \(\Sigma^+\), \(\Sigma^0\), \(\Xi^-\), and \(\Xi^0\). The normalized Discord\(D\) is represented by blue dot-dashed curves, the normalized steering \(S\) by yellow dashed curves, the normalized entanglement \(E\) by black solid curves, the coherence \(C_{l_1}\) by green dotted curves and the concurrence \(C\) by pink dotted curves. In the upper panel, the Ohmicity is fixed at \(s=1\) with t=0.2 and T=0.2.   }
    \label{fig9}
\end{figure*}

The measurement-dependent classical correlation is then written as
\begin{equation}
J(A:B) \equiv S(\rho_A)-\sum_k p_k\, S(\rho_{A|\Pi_k}),
\end{equation}
which has the same general structure as the mutual information above. Since this quantity depends on the chosen measurement basis, the classical part of the correlations is obtained by maximizing over all projectors $\{\Pi_k\}$. The quantum discord is therefore defined as
\begin{align}
D[\rho_{AB}]
&\equiv I(A:B)-\max_{\{\Pi_k\}} J(A:B)
\nonumber\\[4pt]
&= S(\rho_B)-S(\rho_{AB})
+\min_{\{\Pi_k\}}\sum_k p_k\,S(\rho_{A|\Pi_k}).
\end{align}

From this definition, discord behaves like an entropy-based quantity: it is non-negative and bounded above by unity,
\begin{equation}
0 \le D[\rho_{AB}] \le 1.
\end{equation}
Another important point is that quantum discord is generally asymmetric, since measuring subsystem $A$ instead of $B$ may lead to a different value. In the present work, however, the hyperon-antihyperon states produced in $e^+e^-\to Y\bar{Y}$ are symmetric because of CP symmetry, so the discord is the same whether the measurement is performed on $Y$ or on $\bar{Y}$.The main difficulty in computing quantum discord comes from the optimization over all possible local projective measurements, which makes a closed-form expression unavailable for generic two-qubit states. In our case, the hyperon-antihyperon spin states in $e^+e^-$ scattering are symmetric rank-2 X states with two nonzero eigenvalues
\[
\eta(\tau,T)=e^{-4\Gamma_s(\tau,T)}.
\]

Thus,
\[
\rho_{14}(\tau,T)
=
\frac{
\eta(\tau,T)(t_1-t_2)
}{4},
\qquad
\rho_{23}(\tau,T)
=
\frac{
t_1+t_2
}{4}.
\]

The effective parameters of the dephased $X$ state therefore become
\[
t_1(\tau,T)
=
2\bigl(\rho_{23}+\rho_{14}\bigr)
=
\frac{
(1+\eta)t_1+(1-\eta)t_2
}{2},
\]

\[
t_2(\tau,T)
=
2\bigl(\rho_{23}-\rho_{14}\bigr)
=
\frac{
(1-\eta)t_1+(1+\eta)t_2
}{2},
\]

\[
t_3(\tau,T)=t_3,
\qquad
a(\tau,T)=a.
\]

The eigenvalues of the dephased state are then given by
\[
\lambda_{1,4}(\tau,T)
=
\frac{1}{4}
\left[
1+t_3
\pm
\sqrt{
4a^2+\eta(\tau,T)^2(t_1-t_2)^2
}
\right],
\]

\[
\lambda_{2,3}(\tau,T)
=
\frac{1}{4}
\left(
1-t_3
\pm
(t_1+t_2)
\right).
\]

\begin{figure*}[t] 
         \includegraphics[width=18.8cm]{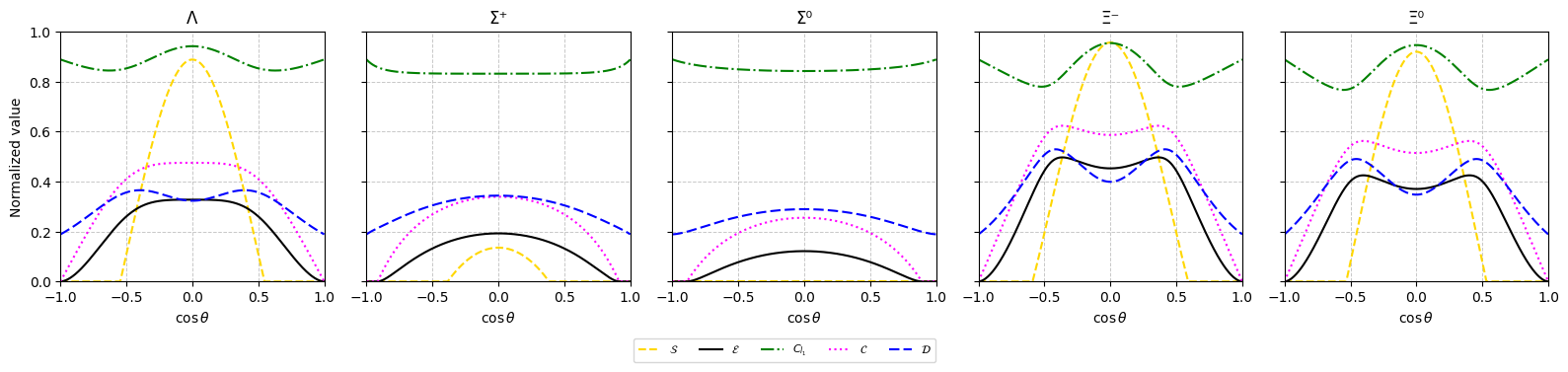}
      \caption{We display several quantum correlation measures as functions of \(\cos\vartheta\), where \(\vartheta\) denotes the scattering angle, for the process \(e^+e^- \to J/\psi \to Y\bar{Y}\). The five panels, labeled from (a) to (e), correspond respectively to \(\Lambda\), \(\Sigma^+\), \(\Sigma^0\), \(\Xi^-\), and \(\Xi^0\). The normalized Discord\(D\) is represented by blue dot-dashed curves, the normalized steering \(S\) by yellow dashed curves, the normalized entanglement \(E\) by black solid curves, the coherence \(C_{l_1}\) by green dotted curves and the concurrence \(C\) by pink dotted curves. In the upper panel, the Ohmicity is fixed at  \(s=2.5\) with t=0.2 and T=0.2.}
    \label{fig10}
\end{figure*}

An important point is that the state under dephasing is not necessarily still a rank-2 state. Therefore, the specific simplification used for the initial state cannot be directly applied anymore. One must instead employ the general analytical formula for the quantum discord of an $X$ state with the dephased parameters. Hence, the quantum discord reads
\begin{equation}
\begin{aligned}
D[\rho^X(\tau,T)] ={}& 1
-\frac{1+a}{2}\log_2\!\left(\frac{1+a}{2}\right)
-\frac{1-a}{2}\log_2\!\left(\frac{1-a}{2}\right) \\
&+\sum_{i=1}^{4}\lambda_i(\tau,T)\log_2\lambda_i(\tau,T)
-\max_{\epsilon\in[0,1]} F(\epsilon;\tau,T).
\end{aligned}
\end{equation}

\begin{equation}
\begin{aligned}
F(\epsilon;\tau,T) ={}& 1
+\frac{1+a\epsilon+H_+}{4}
\log_2\!\left(
\frac{1+a\epsilon+H_+}{1+a\epsilon}
\right) \\
&+\frac{1+a\epsilon-H_+}{4}
\log_2\!\left(
\frac{1+a\epsilon-H_+}{1+a\epsilon}
\right) \\
&+\frac{1-a\epsilon+H_-}{4}
\log_2\!\left(
\frac{1-a\epsilon+H_-}{1-a\epsilon}
\right) \\
&+\frac{1-a\epsilon-H_-}{4}
\log_2\!\left(
\frac{1-a\epsilon-H_-}{1-a\epsilon}
\right).
\end{aligned}
\end{equation}

with
\[
H_\pm(\epsilon;\tau,T)
=
\sqrt{
t(\tau,T)^2(1-\epsilon^2)
+
(a\pm t_3\epsilon)^2
},
\]

and
\[
t(\tau,T)
=
\max\left\{
|t_1(\tau,T)|,\,
|t_2(\tau,T)|
\right\}.
\]
is the binary Shannon entropy. The quantities $a$ and $t_1,t_2,t_3$ are functions of $\vartheta$ and depend on the parameters $\alpha_\psi$ and $\Delta\Phi$.For simplicity, we denote the quantum discord of the hyperon-antihyperon state by $D \equiv D[\rho_{Y\bar{Y}}]$. In the particular case where the optimum is attained at $\epsilon=0$, the quantum discord reduces to
\begin{equation}
D\left[ \rho_{Y\bar{Y}}(\tau,T)\right] =h\left(\frac{1+a}{2}\right)-h\left(\frac{1+t_3}{2}\right)+h\left(\frac{1+t(\tau,T)}{2}\right),
\end{equation}

where $h(x)=-x\log_2 x-(1-x)\log_2(1-x)$ is the binary Shannon entropy, and
\begin{equation}
\begin{aligned}
t(\tau,T)
={}&
\max\Bigg\{
\left|
\frac{
(t_1+t_2)
+e^{-4\Gamma_s(\tau,T)}(t_1-t_2)
}{2}
\right|,
\\[2mm]
&
\left|
\frac{
(t_1+t_2)
-e^{-4\Gamma_s(\tau,T)}(t_1-t_2)
}{2}
\right|
\Bigg\}.
\end{aligned}
\end{equation}
Entanglement is often quantified by the concurrence; however, in this work, we instead employ the entanglement of formation as a more informative measure. It is defined as:
\begin{equation}
E \equiv 
h\!\left(
\frac{1+\sqrt{1-C^2[\rho_{Y\bar{Y}}]}}{2}
\right)
=
h\!\left(
\frac{1+\sqrt{1-t_2^2}}{2}
\right),
\end{equation}
where $h(x)$ denotes the binary Shannon entropy. In the second equality, we have used the relation $C[\rho_{Y\bar{Y}}]=|t_2|$. By construction, $E$ behaves as an entropy-like quantity and satisfies $E\in[0,1]$. In close analogy with Bell nonlocality, quantum steering is certified through the violation of steering inequalities. When such an inequality is violated, the underlying state cannot be explained by a local-hidden-state model. For two-qubit systems, Cavalcanti, Jones, Wiseman, and Reid introduced a steering inequality with three measurement settings on each side, namely

\[
F^{\mathrm{CJWR}}_{3}
\equiv
\frac{1}{\sqrt{3}}
\left|
\sum_{k=1}^{3}
\operatorname{Tr}\!\left[\rho\,(A_k\otimes B_k)\right]
\right|
\le 1,
\]
\[
\text{where } A_k=\mathbf{s}_k\cdot\boldsymbol{\sigma},\qquad
B_k=\mathbf{r}_k\cdot\boldsymbol{\sigma},
\]

$\mathbf{s}_k$ and $\mathbf{r}_k$ ($k=1,2,3$) are unit vectors, and $\{\mathbf{r}_1,\mathbf{r}_2,\mathbf{r}_3\}$ forms an orthogonal Cartesian basis. The steering expression reaches its maximum when the measurement directions are chosen such that

\[
\|\mathcal{C}\mathbf{r}_1\|
=
\|\mathcal{C}\mathbf{r}_2\|
=
\|\mathcal{C}\mathbf{r}_3\|
=
\sqrt{\frac{\operatorname{Tr}(\mathcal{C}^{T}\mathcal{C})}{3}},
\qquad
\mathbf{s}_k
=
\frac{\mathcal{C}\mathbf{r}_k}{\sqrt{\operatorname{Tr}(\mathcal{C}^{T}\mathcal{C})/3}},
\]
$\mathcal{C}_{ij}$ denotes the  $3\times 3$ correlation matrix. Consequently, the maximal CJWR violation is
\[
F_3[\rho]
\equiv
\max_{\mathbf{s}_k,\mathbf{r}_k}
F^{\mathrm{CJWR}}_{3}
=
\sqrt{\operatorname{Tr}(\mathcal{C}^{T}\mathcal{C})}.
\]

Bell nonlocality and quantum steering share a common feature: both are identified through the violation of specific inequalities, namely Bell inequalities and CJWR inequalities, respectively. For quantum steering, the relevant threshold is $F_3>1$, with a maximal value at t=0 is given by $F_3^{\max}=\sqrt{3}$. A normalized measure of steering can thus be introduced as
\begin{equation}
S \equiv 
\max\left\{
0,\,
\frac{F_3-1}{Max(F_3)-1}
\right\}
\in [0,1].
\end{equation}

We therefore have
\[
F_3\!\left[\rho_{Y\bar Y}(\tau,T)\right]
= C_{11}^2 + C_{22}^2 + C_{33}^2.
\]We therefore have
\[
F_3\!\left[\rho_{Y\bar Y}(\tau,T)\right]
=
\sqrt{C_{11}^2 + C_{22}^2 + C_{33}^2}.
\]
By substituting $C_{11}$, $C_{22}$, and $C_{33}$, one obtains

\begin{equation}
\begin{aligned}
F_3\!\left[\rho_{Y\bar Y}(\tau,T)\right]
={}&
\sqrt{
\frac14\left(
(1+\eta)t_1 + (1-\eta)t_2
\right)^2
}\\
&{\quad+
\frac14\left(
(1-\eta)t_1 + (1+\eta)t_2
\right)^2
+t_3^2
}.
\end{aligned}
\end{equation}

Expanding the two squares gives
\begin{equation}
\begin{aligned}
\left(
(1+\eta)t_1 + (1-\eta)t_2
\right)^2
={}&
(1+\eta)^2 t_1^2
\\
&+2(1-\eta^2)t_1 t_2
+(1-\eta)^2 t_2^2 ,
\end{aligned}
\end{equation}

\begin{equation}
\begin{aligned}
\left(
(1-\eta)t_1 + (1+\eta)t_2
\right)^2
={}&
(1-\eta)^2 t_1^2
\\
&+2(1-\eta^2)t_1 t_2
+(1+\eta)^2 t_2^2 .
\end{aligned}
\end{equation}
Adding these two expressions yields
\begin{equation}
\begin{aligned}
&
\left(
(1+\eta)t_1 + (1-\eta)t_2
\right)^2
\\
&\quad+
\left(
(1-\eta)t_1 + (1+\eta)t_2
\right)^2
\\[2mm]
&=
\Big[
(1+\eta)^2 + (1-\eta)^2
\Big]
(t_1^2+t_2^2)
\\
&\quad+
4(1-\eta^2)t_1 t_2 .
\end{aligned}
\end{equation}

Since
\[
(1+\eta)^2 + (1-\eta)^2 = 2(1+\eta^2),
\]
we can factorize the result as
\[
C_{11}^2 + C_{22}^2
=
\frac12\Big[(t_1+t_2)^2 + \eta^2 (t_1-t_2)^2\Big].
\]

Because $\eta = e^{-4\Gamma_s(\tau,T)}$, thus
\[
F_3\!\left[\rho_{Y\bar Y}(\tau,T)\right]
=
\sqrt{
t_3^2
+
\frac12(t_1+t_2)^2
+
\frac12 e^{-8\Gamma_s(\tau,T)}(t_1-t_2)^2
}.
\]

The correct form of the quantity $S(\tau,T)$ is
\[
S(\tau,T)\equiv \max\left\{0,\frac{3}{F_3\!\left[\rho_{Y\bar Y}(\tau,T)\right]}-1\right\}\in[0,1],
\]
we obtain directly
{\small \begin{align}
 S(\tau,T)
=
\max\left\{
0,\,
\frac{3}{
t_3^2
+
2(t_1+t_2)^2
+
e^{-8\Gamma_s(\tau,T)}(t_1-t_2)^2
}
-1
\right\}.   
\end{align}}
The concurrence is a standard monotone entanglement and is closely related to the entanglement of formation. In the present work, we use concurrence to characterize the entanglement of the hyperon-antihyperon state.

For a general two-qubit state, the concurrence is defined as
\begin{equation}
C = \max\left(0,\lambda_1-\lambda_2-\lambda_3-\lambda_4\right),
\end{equation}
where $\lambda_i$ $(i=1,2,3,4)$ are the eigenvalues, in decreasing order, of the matrix
\begin{equation}
\rho\,(\sigma_y \otimes \sigma_y)\,\rho^*\,
(\sigma_y \otimes \sigma_y).
\end{equation}

For an $X$ state, the Wootters concurrence is given by
\[
C(\rho)=2\max\left\{
0,\,
|\rho_{14}|-\sqrt{\rho_{22}\rho_{33}},\,
|\rho_{23}|-\sqrt{\rho_{11}\rho_{44}}
\right\}.
\]
By substituting the elements of
$\rho_{Y\bar Y}^{X}(\tau,T)$, we obtain
\begin{equation}
\begin{aligned}
C(\tau,T)
=2\max\Bigg\{&
0,
\frac{
e^{-4\Gamma_s(\tau,T)}
|t_1-t_2|
}{4}
-\frac{1-t_3}{4},
\\[2mm]
&
\frac{|t_1+t_2|}{4}
-\frac{
\sqrt{
(1+2a+t_3)(1-2a+t_3)
}
}{4}
\Bigg\}.
\end{aligned}
\end{equation}

Equivalently, this can be written as
\begin{equation}
\begin{aligned}
C(\tau,T)
=\max\Bigg\{&
0,
\frac{
e^{-4\Gamma_s(\tau,T)}
|t_1-t_2|
-(1-t_3)
}{2},
\\[2mm]
&
\frac{
|t_1+t_2|
-\sqrt{(1+t_3)^2-4a^2}
}{2}
\Bigg\}.
\end{aligned}
\end{equation}

In the present case, however, the state has rank 2 and exhibits a particular symmetry: only two eigenvalues are nonzero, while the remaining two vanish. Moreover, the parameters $a$, $t_1$, $t_2$, and $t_3$ are not independent, but are constrained by their explicit expressions in terms of the physical quantities $\alpha_\psi$, $\beta_\psi$, and $\theta$. In Fig.\ref{fig9}-\ref{fig10}, we compare four types of quantum correlations; Bell nonlocality $B$, steering $S$, entanglement $E$, and discord $D$, as functions of $\cos\vartheta$ for several hyperon-antihyperon channels in
$e^+e^- \to J/\psi \to Y\bar{Y}$. The panels  correspond to $\Lambda$, $\Sigma^+$, $\Sigma^0$, $\Xi^-$, and $\Xi^0$, respectively. In each case, the curves for $B$, $S$, $E$, and $D$ are shown as blue dash-dotted, yellow dashed, black solid, and red dotted lines. The results reveal a clear ordering among the first three measures, namely $B < S < E$. Where the discord does not follow a simple hierarchy with respect to the other quantities. Although both discord and entanglement are entropy-like measures, their magnitudes are not directly ordered in general~\cite{61}, showing that they capture different aspects of nonclassical correlation. Beyond their absolute values, the angular regions where these correlations remain nonzero also provide useful information. Where both discord and entanglement are nonvanishing over the full angular range, except at the two collinear limits $\vartheta=0^\circ$ and $180^\circ$. Steering is more restricted and survives only in a narrower region around the transverse direction. Interestingly, increasing the Ohmicity parameter from s=1 to s=2.5 in Fig.\ref{fig10} leads to an enhancement of quantum discord, while the quantum Fisher information associated with temperature estimation is reduced. This indicates that the generated quantum correlations are not directly linked to the thermal sensitivity of the probe. In other words, although the system becomes more quantum in terms of correlations, these correlations do not encode useful information about the temperature, resulting in a degradation of the estimation precision. The angular positions at which the maxima and minima occur remain unchanged; however, the Ohmicity primarily affects their amplitudes by weakening the peaks, as a consequence of the degradation of quantum correlations. Where raising the Ohmicity parameter weakens the quantum correlations that are most sensitive to environmental decoherence. Physically, a larger value of $s$ reshapes the bath spectrum and reduces the survival of the off-diagonal coherences that support entanglement and steering. Since steering requires stronger and more directional correlations than entanglement, it is the first resource to be significantly suppressed. For $\Sigma$ and $\Sigma^{+}$ at $s=3$, the complete suppression of steering does not imply a deterioration of thermal estimation precision. This indicates that steerability and thermometric usefulness are not monotonically related in the present open-system dynamics. While the environment destroys the nonlocal correlations responsible for CJWR violation, it may simultaneously enhance the temperature susceptibility of the evolved state, thereby increasing the quantum Fisher information. Hence, the metrological performance is governed not by steering alone, but by the overall temperature response of the reduced density matrix. The concurrence follows the same trend because it directly tracks the loss of inseparability in the reduced two-qubit state. By contrast, quantum discord may increase because it captures a broader notion of nonclassical correlation, including measurement-induced disturbance in mixed states. Therefore, the environment can simultaneously destroy nonlocal correlations while enhancing weaker quantum correlations that remain visible beyond entanglement.

\section{Conclusion}

In this work, we investigated single-qubit quantum thermometry based on dephasing and showed that it can serve as an efficient mechanism for estimating the temperature of Ohmic samples. The proposed protocol is genuinely quantum in nature, as it relies on the qubit's sensitivity to decoherence and does not require the probe to thermalize with the system under study. The analysis of the individual variance in temperature estimation within Ohmic-type reservoirs provides a clear framework for identifying optimal parameter regimes that minimize estimation errors. By systematically exploring the roles of the ohmicity parameter $  s  $, the deviation angle $  \theta  $, and the decay parameters $  \alpha  $ and $  \beta  $ \cite{ref41}, we determine the most favorable conditions for accurate quantum thermometry. Our protocol employs single-qubit probes coupled to structured thermal reservoirs belonging to the Ohmic family in an exactly solvable pure-dephasing model. The variance depends strongly on temperature in the low-$  T  $ regime but is significantly improved by increasing the number of measurements, confirming the suitability of this approach for quantum sensing at low temperatures. At high temperatures, thermal agitation dominates, leading to saturation and reduced sensitivity to environmental details. The QFI exhibits clear maxima at finite interaction times, particularly pronounced at low temperatures for sub-Ohmic and Ohmic cases, while super-Ohmic environments flatten the peaks and shift optimal performance toward higher temperatures. The quantum signal-to-noise ratio (QSNR) remains small at low $  T  $, increases with temperature, and saturates at high temperatures where spectral details become negligible. Comparative analysis of mutual versus local estimation strategies reveals that, for short times, common-bath configurations with probes such as $  \Sigma^+  $ and $  \Sigma^0  $ yield superior precision thanks to bath-induced correlations \cite{43}. We also observed that attaining larger values of the QFI requires the probe to interact long enough with the sample so that it loses a sufficient amount of coherence and acquires more information about the temperature. However, the optimal regime is not necessarily the one of strongest decoherence, since in that limit the qubit may retain too little information to be useful for estimation. Our results therefore show that the optimal sensing conditions arise from a subtle balance between the dephasing dynamics and the specific Ohmic character of the environment. In addition, we showed that non-Markovian effects do not contribute to improving the estimation protocol, because they occur only for a probe coupled to a super-Ohmic bath at low temperature and do not lead to any enhancement of precision. For longer times, local baths prove more effective. Overall, negative diffusion coefficients slow decoherence and enhance sensitivity in collective settings for specific hyperons.
These results highlight the existence of finite temporal ($  t_{\rm opt}  $) and thermal ($  T_{\rm opt}  $) optima for temperature estimation. At low temperatures, longer probing times are required due to slower decoherence, while higher ohmicity accelerates coherence loss and shifts optimal. On the other hand, at short interaction times, compensating for the loss generated by the Ohmicity requires higher optimal temperatures $T_{\mathrm{opt}}$.
The interplay between spectral properties, particle characteristics (especially hyperons like $  \Sigma  $ and $  \Sigma^0  $) \cite{ref46,ref48}, and estimation strategy (mutual vs. local) offers valuable insights for designing high-precision quantum sensors, particularly in low-temperature regimes where quantum advantages are most pronounced. These findings open promising avenues for further exploration in quantum metrology and thermal sensing applications. Our analysis of hyperon-antihyperon channels ($  \Lambda  $, $  \Sigma^+  $, $  \Sigma^0  $, $  \Xi^-  $, $  \Xi^0  $) confirms this hierarchy while revealing their distinct angular dependence and sensitivity to the environmental spectral properties (Ohmicity parameter $  s  $). Entanglement and discord prove remarkably robust, persisting over nearly the full angular range, whereas steering and Bell nonlocality are confined to narrower angular windows. The interplay between these correlations and the thermal environment further highlights a subtle trade-off: while increasing Ohmicity tends to suppress steering and entanglement (thereby degrading thermometric precision), it can enhance quantum discord, underscoring that not all forms of quantum correlation are equally useful for quantum sensing tasks.
Taken together, these results provide a comprehensive picture of the quantum resource landscape in hyperon-antihyperon systems and offer valuable guidance for harnessing such correlations in quantum metrology and information protocols.
\section*{ACKNOWLEDGMENTS}

A.H acknowledges the financial support of the National Center for Scientific and Technical Research (CNRST) through the "PhD-Associate Scholarship-PASS" program. The authors acknowledge the LPHE-MS, FSR for the technical support. 

\section*{Declaration of competing interest:} 
The authors declare that they have no known competing financial interests or personal relationships that could have appeared to influence the work reported in this paper.

\section*{Data availability:}
No data was used for the research described in the article.
\appendix
\section{Master equation}
We define the following parameters in order to simplify the expressions:
\begin{equation}
\begin{aligned}
\alpha &= a+b,
&
\beta &= b+d,
\\
\gamma &= a+d,
&
\delta &= ad-x^2,
\\
\varepsilon &= b^2-z^2,
&
\zeta &= a\gamma-x^2 = a^2+\delta,
\\
\eta &= d\gamma-x^2 = d^2+\delta,
&
\theta &= dx,
\\
\iota &= ax,
&
\kappa &= x^2,
\\
\xi &= a+2b+d,
&
\sigma &= 2b^2-z^2 ,
\\[1mm]
\sigma'
&=
\dfrac{\sigma}{4b(b^2-z^2)},
&
\theta'
&=
-\dfrac{\theta}{2\gamma\delta},
\\[3mm]
\kappa'
&=
\dfrac{\kappa}{2\gamma\delta},
&
\eta'
&=
\dfrac{\eta}{2\gamma\delta},
\\[3mm]
\zeta'
&=
\dfrac{\zeta}{2\gamma\delta},
&
\iota'
&=
-\dfrac{\iota}{2\gamma\delta},
\\[3mm]
\tau
&=
z^4(-b^2+z^2)
=
-z^4\varepsilon ,
&
\upsilon
&=
4bz^2(b^2-z^2)
=
4bz^2\varepsilon ,
\\[3mm]
\phi &= xz\,\xi,
&
\chi &= x\nu,
\\
\psi &= z\rho,
&
\omega &= z\mu .
\end{aligned}
\end{equation}
\begin{equation}
\begin{aligned}
\lambda
={}&
ab^2+2abd+ad^2-az^2
+b^3+2b^2d+bd^2
\\
&-bx^2-bz^2-dx^2 ,
\\[2mm]
\mu
={}&
-b^2-2bd-d^2-x^2+z^2 ,
\\
\nu
={}&
-ab-ad-b^2-bd+x^2-z^2 ,
\\[2mm]
\pi
={}&
a^2b+a^2d+2ab^2+2abd
\\
&-ax^2+b^3+b^2d
-bx^2-bz^2-dz^2 ,
\\[2mm]
\rho
={}&
-a^2-2ab-b^2-x^2+z^2 ,
\\[2mm]
\Delta
={}&
a^2b^2 + 2a^2bd + a^2d^2
-a^2z^2
\\
&+ 2ab^3 + 4ab^2d + 2abd^2
\\
&- 2abx^2 - 2abz^2
- 2adx^2
\\
&+ b^4 + 2b^3d + b^2d^2
\\
&- 2b^2x^2 - 2b^2z^2
\\
&- 2bdx^2 - 2bdz^2
\\
&- d^2z^2 + x^4
- 2x^2z^2 + z^4 .
\end{aligned}
\end{equation}


\begin{thebibliography}{40}


\bibitem{1}
C. L. Degen, F. Reinhard, and P. Cappellaro,
``Quantum sensing,''
\textit{Rev. Mod. Phys.} \textbf{89}, 035002 (2017).

\bibitem{2}
M. G. A. Paris,
``Quantum estimation for quantum technology,''
\textit{Int. J. Quantum Inf.} \textbf{7}, 125 (2009).

\bibitem{3}
D. Braun, G. Adesso, F. Benatti, R. Floreanini,
U. Marzolino, M. W. Mitchell, and S. Pirandola,
``Quantum-enhanced measurements without entanglement,''
\textit{Rev. Mod. Phys.} \textbf{90}, 035006 (2018).

\bibitem{4}
J. F. Haase, A. Smirne, S. F. Huelga,
J. Kołodyński, and F. Demkowicz-Dobrzański,
``Precision limits in quantum metrology with open quantum systems,''
\textit{Quantum Meas. Quantum Metr.} \textbf{5}, 13 (2018).

\bibitem{5}
C. Benedetti, F. Buscemi, P. Bordone, and M. G. A. Paris,
``Quantum probes for the spectral properties of a classical environment,''
\textit{Phys. Rev. A} \textbf{89}, 032114 (2014).

\bibitem{6}
G. L. Giorgi, F. Galve, and R. Zambrini,
``Probing the spectral density of a dissipative qubit via quantum synchronization,''
\textit{Phys. Rev. A} \textbf{94}, 052121 (2016).
\bibitem{gx1}
A. Candeloro, S. Razavian, M. Piccolini, B. Teklu, S. Olivares, and M. G. A. Paris,
``Quantum probes for the characterization of nonlinear media,''
\textit{Entropy}, vol. 23, no. 10, p. 1353, 2021.
\bibitem{gx3}
B. Teklu, M. G. Genoni, S. Olivares, and M. G. A. Paris,
``Phase estimation in the presence of phase diffusion: the qubit case,''
\textit{Physica Scripta}, vol. 2010, no. T140, p. 014062, 2010.

\bibitem{9}
A. Usui, B. Buča, and J. Mur-Petit,
``Quantum probe spectroscopy for cold atomic systems,''
\textit{New J. Phys.} \textbf{20}, 103006 (2018).

\bibitem{10}
M. Brunelli, S. Olivares, and M. G. A. Paris,
``Qubit thermometry for micromechanical resonators,''
\textit{Phys. Rev. A} \textbf{84}, 032105 (2011).

\bibitem{11}
L. A. Correa, M. Mehboudi, G. Adesso, and A. Sanpera,
``Individual Quantum Probes for Optimal Thermometry,''
\textit{Phys. Rev. Lett.} \textbf{114}, 220405 (2015).

\bibitem{12}
A. De Pasquale, K. Yuasa, and V. Giovannetti,
``Estimating temperature via sequential measurements,''
\textit{Phys. Rev. A} \textbf{96}, 012316 (2017).

\bibitem{13}
S. Campbell, M. G. Genoni, and S. Deffner,
``Precision thermometry and the quantum speed limit,''
\textit{Quantum Sci. Technol.} \textbf{3}, 025002 (2018).

\bibitem{14}
K. V. Hovhannisyan and L. A. Correa,
``Measuring the temperature of cold many-body quantum systems,''
\textit{Phys. Rev. B} \textbf{98}, 045101 (2018).

\bibitem{15}
M. Mehboudi, A. Sanpera, and L. A. Correa,
``Thermometry in the quantum regime: recent theoretical progress,''
\textit{J. Phys. A: Math. Theor.} \textbf{52}, 303001 (2019).

\bibitem{16}
C. Benedetti, F. S. Sehdaran, M. H. Zandi,
and M. G. A. Paris,
``Quantum probes for the cut-off frequency of Ohmic environments,''
\textit{Phys. Rev. A} \textbf{97}, 012126 (2018).



\bibitem{18}
S. Razavian, C. Benedetti, M. Bina,
Y. Akbari-Kourbolagh, and M. G. A. Paris,
``Quantum thermometry by single-qubit dephasing,''
\textit{Eur. Phys. J. Plus} \textbf{134}, 284 (2019).

\bibitem{19}
F. Salari Sehdaran, M. Bina,
C. Benedetti, and M. G. A. Paris,
``Quantum Probes for Ohmic Environments at Thermal Equilibrium,''
\textit{Entropy} \textbf{21}(5), 486 (2019).

\bibitem{20}
H. Cramér,
\textit{Mathematical Methods of Statistics}
(1962).











\bibitem{40}
J. F. Clauser and M. A. Horne,
``Experimental consequences of objective local theories,''
\textit{Phys. Rev. D} \textbf{10}, 526 (1974).

\bibitem{41}
A. Aspect, P. Grangier, and G. Roger,
``Experimental Tests of Realistic Local Theories via Bell's Theorem,''
\textit{Phys. Rev. Lett.} \textbf{47}, 460 (1981).

\bibitem{42}
A. J. Barr, M. Fabbrichesi, R. Floreanini,
E. Gabrielli, and L. Marzola,
``Quantum information and entanglement in particle physics,''
\textit{Prog. Part. Nucl. Phys.} \textbf{139}, 104134 (2024).

\bibitem{43}
A. Bernal,
``Quantum correlations in particle physics,''
\textit{Phys. Rev. D} \textbf{109}, 116007 (2024).

\bibitem{44}
Y. Afik and J. R. M. de Nova,
``Entanglement and quantum tomography with particle decays,''
\textit{Eur. Phys. J. Plus} \textbf{136}, 907 (2021).

\bibitem{45}
M. Fabbrichesi, R. Floreanini, and G. Panizzo,
``Testing Bell inequalities in high-energy physics,''
\textit{Phys. Rev. Lett.} \textbf{127}, 161801 (2021).

\bibitem{46}
Y. Afik and J. R. M. de Nova,
``Quantum information with hadronic systems,''
\textit{Quantum} \textbf{6}, 820 (2022).

\bibitem{47}
A. Collaboration et al.,
``Observation of quantum entanglement in high-energy physics,''
\textit{Nature} \textbf{633}, 542 (2024).
\bibitem{gx2}
J. Loulijat, A. Slaoui, M. Gouighri, and B. Teklu,
``Influence of quantum decoherence on the survival of quantumness in neutrino oscillations,''
\textit{Scientific Reports}, 2026.

\bibitem{gx4}
K. El Bouzaidi, A. Slaoui, L. B. Drissi, E. H. Saidi, and R. Ahl Laamara,
``Dynamics of quantum information resources in two-flavor neutrino oscillations,''
\textit{The European Physical Journal C}, vol. 85, no. 11, p. 1349, 2025.

\bibitem{48}
M. Fabbrichesi, R. Floreanini, and E. Gabrielli,
``Quantum correlations in particle decays,''
\textit{Eur. Phys. J. C} \textbf{83} (2023).

\bibitem{49}
K. Ehatäht, M. Fabbrichesi,
L. Marzola, and C. Veelken,
``Entanglement studies in particle collisions,''
\textit{Phys. Rev. D} \textbf{109}, 032005 (2024).

\bibitem{50}
T. Han, M. Low, and Y. Su,
``Entanglement and Bell Nonlocality in $\tau^+\tau^-$ at the BEPC,''
arXiv:2501.04801 [hep-ph] (2025).

\bibitem{51}
A. J. Barr,
``Testing Bell inequalities at colliders,''
\textit{Phys. Lett. B} \textbf{825}, 136866 (2022).

\bibitem{52}
A. J. Barr, P. Caban, and J. Rembieliński,
``Quantum entanglement in relativistic systems,''
\textit{Quantum} \textbf{7}, 1070 (2023).

\bibitem{53}
J. A. Aguilar-Saavedra, A. Bernal,
J. A. Casas, and J. M. Moreno,
``Quantum correlations in particle-antiparticle systems,''
\textit{Phys. Rev. D} \textbf{107}, 016012 (2023).

\bibitem{54}
N. A. Törnqvist,
``Possible large baryon-antibaryon enhancements,''
\textit{Found. Phys.} \textbf{11}, 171 (1981).

\bibitem{55}
C. Qian, J.-L. Li, A. S. Khan,
and C.-F. Qiao,
``Entanglement entropy in hyperon systems,''
\textit{Phys. Rev. D} \textbf{101}, 116004 (2020).

\bibitem{56}
S. Wu, C. Qian, Q. Wang,
and X.-R. Zhou,
``Quantum correlations in hyperon-antihyperon systems,''
\textit{Phys. Rev. D} \textbf{110}, 054012 (2024).

\bibitem{57}
M. Fabbrichesi, R. Floreanini,
E. Gabrielli, and L. Marzola,
``Bell inequalities and hyperon decays,''
\textit{Phys. Rev. D} \textbf{110}, 053008 (2024).

\bibitem{58}
J. Pei, X. Hao, X. Wang, and T. Li,
``Observation of quantum entanglement in $\Lambda\bar{\Lambda}$ pair production via electron-positron annihilation,''
arXiv:2505.09931 [hep-ph] (2025).

\bibitem{59}
K. Cheng and B. Yan,
``Quantum nonlocality in baryonic systems,''
\textit{Phys. Rev. Lett.} \textbf{135}, 011902 (2025).

\bibitem{60}
M. Fabbrichesi, R. Floreanini,
E. Gabrielli, and L. Marzola,
``Multipartite quantum correlations in high-energy systems,''
\textit{Phys. Rev. D} \textbf{109}, L031104 (2024).

\bibitem{61}
S. Wu, C. Qian, Y.-G. Yang,
and Q. Wang,
``Generalized quantum measurement in spin-correlated hyperon-antihyperon decays,''
arXiv:2402.16574 [hep-ph] (2024).

\bibitem{62}
M. Ablikim et al.,
``Observation of quantum correlations in baryon systems,''
\textit{Nature Communications} \textbf{16},
10.1038/s41467-025-59498-4 (2025).


\bibitem{ref40}
M.~Ablikim \textit{et al.} (BESIII Collaboration),
Nature Physics \textbf{15}, 631 (2019).

\bibitem{ref41}
M.~Ablikim \textit{et al.} (BESIII Collaboration),
Phys.\ Rev.\ D \textbf{95}, 052003 (2017).

\bibitem{ref42}
M.~Ablikim \textit{et al.} (BES Collaboration),
Phys.\ Rev.\ D \textbf{78}, 092005 (2008).

\bibitem{ref43}
M.~Ablikim \textit{et al.} (BESIII Collaboration),
Phys.\ Rev.\ Lett.\ \textbf{125}, 052004 (2020).

\bibitem{ref44}
M.~Ablikim \textit{et al.} (BESIII Collaboration),
Phys.\ Rev.\ Lett.\ \textbf{133}, 101902 (2024).

\bibitem{ref45}
M.~Ablikim \textit{et al.} (BESIII Collaboration),
Nature \textbf{606}, 64 (2022).

\bibitem{ref46}
R.~L.~Workman \textit{et al.} (Particle Data Group),
Prog.\ Theor.\ Exp.\ Phys.\ \textbf{2022}, 083C01 (2022).

\bibitem{ref47}
M.~Ablikim \textit{et al.},
Phys.\ Lett.\ B \textbf{770}, 217 (2017).

\bibitem{ref48}
M.~Ablikim \textit{et al.} (BESIII Collaboration),
Phys.\ Rev.\ D \textbf{108}, L031106 (2023).

\end{thebibliography}
\end{document}